\documentclass[twocolumn]{aastex631}

\newcommand\lya{{\rm\,Ly$\alpha$}}
\newcommand\ha{{\rm\,H$\alpha$}}
\newcommand\hb{{\rm\,H$\beta$}}
\newcommand\oiii{{\rm\,[O{\sc iii}]}}

\newcommand\hii{{\rm\,H{\sc ii}}}

\newcommand\Ms{$\mathrm{M_{\odot}}$}
\usepackage{physics}
\usepackage{color}

\usepackage{booktabs}
\usepackage{ulem}

\newcommand{\simgt}{\,\rlap{\lower 3.5 pt \hbox{$\mathchar \sim$}} \raise
1pt \hbox {$>$}\,}
\newcommand{\simlt}{\,\rlap{\lower 3.5 pt \hbox{$\mathchar \sim$}} \raise
1pt \hbox {$<$}\,}

\graphicspath{{./}{figures/}}

\accepted{2025 June 3}

\shorttitle{High Equivalent Width Emitters at $z\sim7$}
\shortauthors{K. Daikuhara et al.}

\begin{document}

\title{Nature of High Equivalent Width Emitters in the Epoch of Reionization Revealed by JWST Medium-band Imaging}

\correspondingauthor{Kazuki Daikuhara}
\email{daikuhara@astr.tohoku.ac.jp}

\author[0000-0002-9509-2774]{Kazuki Daikuhara}
\affil{Institute of Space and Astronautical Science, Japan Aerospace Exploration Agency, 3-1-1, Yoshinodai, Chuou-ku, Sagamihara, Kanagawa 252-5210, Japan}
\affil{Astronomical Institute, Tohoku University, 6-3, Aramaki, Aoba, Sendai, Miyagi, 980-8578, Japan}

\author[0000-0002-8512-1404]{Takahiro Morishita}
\affiliation{IPAC, California Institute of Technology, MC 314-6, 1200 E. California Boulevard, Pasadena, CA 91125, USA}
\author[0000-0002-2993-1576]{Tadayuki Kodama}
\affil{Astronomical Institute, Tohoku University, 6-3, Aramaki, Aoba, Sendai, Miyagi, 980-8578, Japan}
\author[0000-0001-7583-0621]{Ranga-Ram Chary}
\affiliation{IPAC, California Institute of Technology, MC 314-6, 1200 E. California Boulevard, Pasadena, CA 91125, USA}
\author[0000-0002-2651-1701]{Masayuki Akiyama}
\affil{Astronomical Institute, Tohoku University, 6-3, Aramaki, Aoba, Sendai, Miyagi, 980-8578, Japan}
\author[0000-0002-5963-6850]{Jose. M. Pérez-Martínez}
\affiliation{Instituto de Astrofísica de Canarias (IAC), E-38205, La Laguna, Tenerife, Spain}
\affiliation{Departamento Astrofísica, Universidad de La Laguna, E-38206 La Laguna, Tenerife, Spain}



\begin{abstract}
Extreme emission line galaxies (EELGs) at high redshifts are considered key contributors to cosmic reionization at $z>6$ due to their higher ionization efficiencies.
We have identified 119 \hb+\oiii\ emitters at $z\sim7$ selected by a flux excess in the medium-band filter F410M in the public James Webb Space Telescope Cycle-1 fields.
Our emitters exhibit a wide range in rest-frame \hb+\oiii\ equivalent width (EWs), 420 $<$ EW$_{0}$ / \AA\ $<$ 6850 (with the median value of $\sim1700$\,\AA). 
Among them, 19 are EW$_{0}$ $>$ 3000 / \AA, which represent extreme populations even in the context of recent findings with JWST. 
They are characterized by (i)~low stellar mass ($\sim 3\times10^{7}$\,\Ms), (ii)~blue colors ($\beta_{\rm UV}\sim -2.2$), and (iii)~low dust attenuation ($A_{\mathrm{V}}\sim 0.1$\,mag).
We discuss the physical mechanisms responsible for the observed high rest-frame \hb+\oiii\ EWs, including (1) photoionization by AGN, (2) stellar photoionization in the vicinity of \hii\ regions, and (3) radiative shocks powered by outflows either from AGN or massive stars.
Notably, we find 13 emitters with spatially offset  \hb+\oiii\ emission compared to the UV and stellar components. Given the absence of obvious signatures of actively accreting black holes, these emitters are likely under strong feedback-driven winds from massive stars. Lastly, we report a unique overdensity of EELGs in one of the observed fields. The discovery of such a ``star-bursting"  overdensity supports the idea that large ionizing bubbles formed around some EEGLs in the early Universe.
\end{abstract}



\section{Introduction} 
\label{sec:intro}

Star-forming galaxies (SFGs), some with extreme emission lines \citep[e.g.,][]{vanderWel2011,Atek2011}, are believed to play a pivotal role in the reionization of the intergalactic medium at high redshift. Over the past decade, deep observations by the Hubble Space Telescope (HST) and Spitzer Space Telescope have identified low-mass, SFGs with high equivalent widths (EWs) in the early Universe \citep[e.g.,][]{Chary2005, Shim2011, Atek2011,Roberts-Borsani2016,Maseda2018,Barro2019,DeBarros2019,Mainali2020,Endsley2021,Stefanon2022}

In general, extreme emission line galaxies (EELGs) are compact objects characterized by high EWs in emission lines such as \ha, \oiii, and \lya, indicative of intense ongoing star formation activity.
\citet{Cohn2018} reported that EELGs on average formed $\sim15$\,\% of their total stellar mass within the past 50\,Myr, significantly larger compared to only $\sim4$\% for typical SFGs.
EELGs are also expected to exhibit higher ionization efficiencies and higher escape fractions of ionizing photons than typical SFGs \citep[e.g.,][]{Maseda2020, Endsley2023a, Llerena2024, Boyett2024, Simmonds2024, Simmonds2024b, Caputi2024, Rinaldi2023, Rinaldi2024}.
Notably, EELGs with intense star-forming activity can release Lyman continuum (LyC) photons into the intergalactic medium (IGM) \citep{Llerena2024}.
These escaping photons play a crucial role during the cosmic reionization epoch.
The reionization process is generally understood to result from the cumulative emission of ionizing photons from luminous sources—primarily young SFGs, and possibly also from quasars.
As these sources formed and evolved, they emitted high-energy UV photons that escaped into the IGM, gradually ionizing the surrounding hydrogen gas.
Thus, understanding how ionizing photons escape into the IGM is critical to unraveling the process of cosmic reionization.

The total budget of ionizing photons that governs reionization is determined by the relative abundance of sources, their intrinsic Lyman continuum photon production efficiencies ($\xi_{\mathrm{ion}}$), and the fraction of photons that escape into the IGM ($f_{\mathrm{esc}}$).
Various models have investigated how $f_{\mathrm{esc}}$ and $\xi_{\mathrm{ion}}$ depend on galaxy properties.
\citet{Finkelstein2019} found that low-mass galaxies with high $f_{\mathrm{esc}}$ lead to a more extended reionization history, requiring contributions from quasars at later times.
\citet{Naidu2020} modeled $f_{\mathrm{esc}}$ as a function of star formation rate (SFR) surface density and showed that galaxies with $M_{\mathrm{UV}} < -18$ contribute more than 80\% of the ionizing photons required for reionization.
However, reionization models often face degeneracies, necessitating a comprehensive, multifaceted approach \citep{Greig2017}.
To address this, \citet{Mason2019} employed a Bayesian framework to infer a nonparametric form of the ionizing photon production rate ($\dot{N}{\mathrm{ion}}$) at $z > 6$, demonstrating a rapid decline in $\dot{N}{\mathrm{ion}}$ based on constraints from the cosmic microwave background and \lya\ damping, suggesting that reionization was actively progressing at that epoch.
Despite these advances, the individual contributions of galaxies and active galactic nuclei (AGNs) to reionization remain poorly constrained, both theoretically and observationally.
To understand their respective roles, it is essential to investigate the properties of ionizing sources—such as SFGs, AGN, and quasars—and the physical mechanisms that facilitate the escape of ionizing photons.

Galaxies that have high EWs were also observed in the local Universe as Green Pea galaxies \citep{Cardamone2009,Amorin2010,Izotov2011}.
Green Pea galaxies are low-mass systems in low-density environments, with high SFRs, low metallicities, compact size, and low dust contents \citep{Cardamone2009}.
Since Green Pea galaxies have similar properties to high-$z$ low-mass galaxies (e.g., \lya\ emitters), they are thought to be a local analog of high-$z$ low-mass galaxies \citep[e.g.,][]{Kim2021}. 
The fraction of galaxies showing extreme emission line strengths increases with redshift: from $<0.1$\,\% in the local Universe \citep{Shim2013, Fumagalli2012} to approximately 70\,\% at $z\sim5$ \citep{Shim2011}. 
From $z$ $\sim$ 2 to $z$ $\sim$ 7, the increase is about a factor of 10 \citep[e.g.,][]{Boyett2022}. 
Hence, these EELGs comprise a critical population in the early Universe.

However, the physical conditions that enable galaxies to achieve such high EWs are still not well understood.
Various scenarios have been discussed, including AGNs \citep{Shim2011}, continuous and extreme, bursty star-formation histories \citep{Shim2011, Hopkins2014,Cohn2018}, a top-heavy stellar initial mass function \citep{Chary2008}, and merger-triggered intense starbursts \citep{Gupta2023b}. 
Hydrodynamical simulations can reproduce some EELG populations, but not the most extreme sources, as well as their comoving number density \citep{Wilkins2023}.

\begin{figure*}
\begin{center} 
\includegraphics[width=170mm]{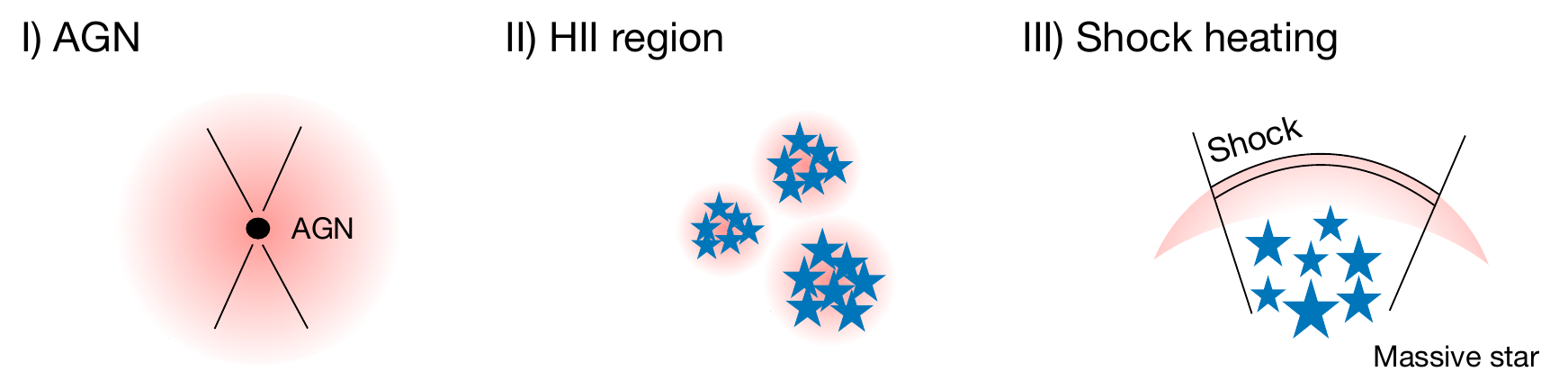}
\end{center} 
\caption{Possible scenarios for the ionizing sources in high-EW emission-line galaxies. High EWs can be caused by photoionization from AGN and \hii\ regions,  and/or shocks from AGN and massive stars. Emission powered by star formation is likely to be more extended than that of AGN. By leveraging the sensitivity and resolution of medium-band filters, we can efficiently assess which physical mechanism is the main contributor by spatially characterizing the \hb\ and \oiii\ emission lines.}
\label{fig:fig}
\end{figure*}

A detailed characterization of EELGs at high redshifts has been severely limited so far, owing to the wavelength coverage of the HST and the limited spatial resolution of Spitzer. Leveraging the unprecedented sensitivity of imaging and spectroscopy provided by JWST, we are now able to examine potential physical mechanisms (see Figure~\ref{fig:fig} for an example).
Early JWST observations have successfully pinpointed EELGs at $z$ = 7 -- 10 \citep{Bradley2023,Endsley2023a,Gupta2023b,Williams2023,Boyett2024,Rinaldi2023,Rinaldi2024}.
In this study, we investigate the nature of SFGs with high EWs by using unique medium-band datasets obtained in the JWST Cycle 1. 
We define high-EW galaxies as those with rest-frame \hb\ + \oiii\ EW $>$ 3000 \AA\ and EELGs as galaxies with rest-frame \hb\ + \oiii\ EW $>$ 1000\AA. 
Although EELGs vary across the literature, we adopt a threshold of \hb\ + \oiii\ EW $=$ 1000\AA. To highlight a particularly extreme subset, we further define galaxies with rest-frame \hb\ + \oiii\ EW $>$ 3000 \AA\ as a distinct population. 
Even within the context of EELG studies, high-EW galaxies are rare objects and a less-explored subset.
JWST's high spatial resolution enables the characterization of the morphology in detail, a key to the origin of high EW emissions. 
This knowledge can provide valuable information for future follow-up spectral observations aimed at discerning kinematics and chemical properties. 
Further, we examine an overdensity of SFGs with high-EW and discuss how this environment contributes to the reionization of EELGs and the Universe.

In Section~\ref{sec:data}, we discuss the imaging data used in this analysis. Section~\ref{sec:method} elaborates on SFG selection and the derivation of their physical properties. The statistical properties of high-EW emitters are covered in Section~\ref{sec:results}. Section~\ref{sec:discussion} and \ref{sec:dis_overdensity} presents discussions on: (i) the nature of ionizing sources causing high-EW, and the origin of star-forming activity, and (ii) the over-densities of EELGs. Lastly, in Section~\ref{sec:conclusion}, we provide a summary of this study. 
We adopt the cosmological parameters of $\mathrm{H_0}$ = 70 km s$^{-1}$ Mpc$^{-1}$, $\Omega_{\mathrm{M}}$ = 0.3, and $\Omega_{\Lambda}$ = 0.7.
Stellar masses ($M_{\star}$) and SFRs are estimated assuming the Chabrier initial mass function \citep[IMF;][]{Chabrier2003}, with a lower mass cutoff of 0.08 $\mathrm{M_{\odot}}$.
All magnitudes in this paper comply with the AB system \citep{Oke1983,Fukugita1996}.

\section{Data} 
\label{sec:data}

\begin{table*}[t]
\begin{center}
\caption{Photometric 5$\sigma$ limiting magnitudes are shown. These values were estimated with a random 0.32 arcsec diameter aperture photometry in empty regions.}

\begin{tabular}{c|cccccccc}
\hline
\quad Field ID\quad & \quad F090W\quad &\quad F115W\quad &\quad F150W\quad &\quad F200W\quad &\quad F277W\quad &\quad F356W\quad &\quad F410M\quad &\quad F444W\quad \\\hline\hline
JADES-GOODS &\quad 29.6\quad &\quad 29.9\quad &\quad 29.8\quad &\quad 29.9\quad &\quad 30.2\quad &\quad 30.1\quad &\quad 29.6\quad &\quad 29.8\quad \\
PRIMER-COSMOS &\quad 27.8\quad &\quad 27.8\quad &\quad 28.0\quad &\quad 28.1\quad &\quad 28.4\quad &\quad 28.5\quad &\quad 27.8\quad &\quad  28.2\quad \\
PRIMER-UDS &\quad 27.8\quad &\quad 27.8\quad &\quad 28.0\quad &\quad 28.1\quad &\quad 28.4\quad &\quad 28.4\quad &\quad 27.6\quad &\quad 27.9\quad \\
NEP &\quad 28.4\quad &\quad 28.5\quad &\quad 28.6\quad &\quad 28.7\quad &\quad 29.0\quad &\quad 29.2\quad &\quad 28.4\quad &\quad  28.6\quad \\
PAR1199 &\quad 29.0 &\quad 29.0\quad &\quad 29.0\quad &\quad 29.2\quad &\quad 29.4\quad &\quad 29.5\quad &\quad 28.9\quad &\quad 28.9\quad \\\hline
\end{tabular}
\label{tab:limmag}
\end{center}
\end{table*}

We utilized a suite of public deep imaging data procured in several programs of JWST Cycle 1. Our fields include CEERS  \citep{Bagley2023,Finkelstein2023}, the Public Release Imaging for Extragalactic Research (PRIMER; PID 1837, Dunlop et al.), the PAR1199 field (11:49:47.31, +22:29:32.1) taken as part of the GTO1199 program \citep{Stiavelli2023}, the North Ecliptic Pole Time-domain field \citep[17:22:47.896, +65:49:21.54;][]{Jansen2018,Windhorst2023}, and JWST Advanced Deep Extragalactic Survey \citep[JADES;][]{Robertson2023,Tacchella2023,Eisenstein2023}.
We used the same dataset presented in \cite{Morishita2023}.

We derived the photometric catalog in each field by following the procedure presented in \citet{Morishita2023}.
Sourced detections and aperture photometries were performed with a $0^{\prime\prime}.32$ diameter aperture, applying the double-imaging mode of {\tt Sextractor} \citep{Bertin1996}.
Photometric 5 sigma limiting magnitude are shown in Table~\ref{tab:limmag}.
Detection images were generated by stacking the F277W, F356W, and F444W filters for each field.
{\tt Sextractor} parameters are shown in Table~\ref{table:Sextractor}.

Total fluxes were deduced using the conversion factor, defined as 
\begin{equation}
C = f_{\mathrm{auto}}/f_{\mathrm{aper}},
\end{equation}
which refers to the ratio of FLUX\_AUTO of {\tt Sextractor} output to the aperture flux. 

\begin{table}
\centering
 \caption{Sextractor parameters for source extractions.}
 \label{table:Sextractor}
  \begin{tabular}{lcc}\hline
  Parameters & \\\hline\hline
  \quad DETECT\_MINAREA  &\quad&  9 \\
  \quad DETECT\_THRESH   &\quad&  1.5\\
  \quad ANALYSIS\_THRESH &\quad&  1.0\\
  \quad DEBLEN\_MINCONT  &\quad&  0.0001 \\
  \quad BACK\_SIZE       &\quad&  128\\
  \quad BACK\_FILTERSIZE &\quad&  5\\
  \quad BACKPHOTO\_TYPE  &\quad&  LOCAL\\
  \quad BACKPHOTO\_THICK &\quad&  20\\
  \quad convolution kernel&\quad&gauss\_2.5\_5$\times$5.conv\\\hline
  \end{tabular}
\end{table}

\section{Method}  
\label{sec:method}
\subsection{Selection of \hb+\oiii\ Emitters at $z\sim7$} \label{subsec:selection}


\begin{figure}
\begin{center} 
\includegraphics[width=82mm]{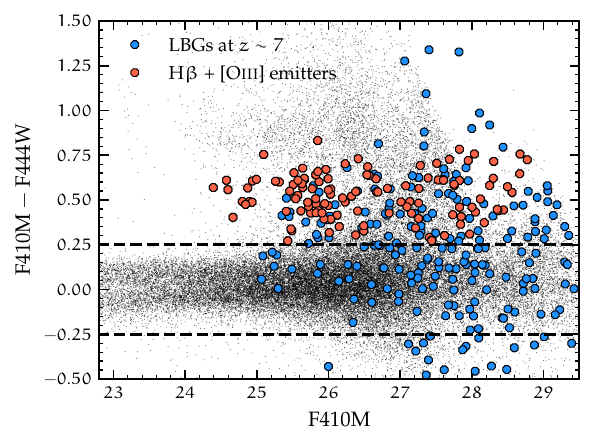}
\end{center} 
\caption{
The color--magnitude diagram. The red circles represent the \hb + \oiii\  emitters at $z\sim 7$, the blue circles represent the \oiii\ emitters, and the black do represent detected sources with SN$_{\mathrm{F444W}}>5$ and SN$_{\mathrm{F410M}}>5$. 
The color-magnitude diagram is utilized to select the objects that show an excess in F410M flux concerning the F444W flux. The solid line shows the EW cut of emitters to avoid including contamination, corresponding to rest-frame \hb + \oiii\  EW$_0$ = 600 \AA.}
\label{fig:selection}
\end{figure}


We select high-$z$ sources using photometric filters to identify the Lyman-break feature \citep{Steidel1998}.
Utilizing a modified version of the F090W dropout selection presented in \cite{Morishita2023}, we securely select \hb+\oiii\ emitters from the dropout candidates. 
Sources that satisfy SN$_{\mathrm{F150W}}$ $>$ 5 detection and SN$_{\mathrm{F090W}}$ $<$ 2 nondetection in the $0^{\prime\prime}.32$ diameter aperture are defined.

We further remove potential contaminants using photometric redshifts and their probability distributions, which include those with potential flux detection in the dropout band (i.e. F090W). Our samples are confined to 7 $<$ $z_{\mathrm{photo}}$ $<$ 8 and $p\,(z > 6)$ $>$ 0.8 \citep[see in detail][]{Morishita2023}. 

After the initial Lyman-break galaxy selection, emitter candidates are then down-selected using the color excess in F410M$-$F444W, satisfying the conditions:
\begin{equation}
  m_{\mathrm{F410M}}-m_{\mathrm{F444W}}>-2.5\log_{10}\left[1-\frac{\,3\sqrt{\sigma_{\mathrm{F410M}}^2+\sigma^2_{\mathrm{F444W}}}}{f_{\mathrm{F410M}}}\right]
\label{eq:SNcut}
\end{equation}
and
\begin{equation}
  m_{\mathrm{F444W}}-m_{\mathrm{F410M}}>0.25,
\label{eq:EWcut}
\end{equation}
where $m$ represents magnitude, $f$ stands for flux density, and $\sigma$ is the photometric flux error.
Equation~(\ref{eq:SNcut}) corresponds to the $3\sigma$ cut of color excess in $m_{\mathrm{F444W}}-m_{\mathrm{F410M}}$.
Equation~(\ref{eq:EWcut}) removes contaminations due to scattering by photometric errors and/or intrinsic colors of objects in $m_{\mathrm{F444W}}-m_{\mathrm{F410M}}$.
Most galaxies on the bright side in F410M magnitude have the color of $m_{\mathrm{F444W}}-m_{\mathrm{F410M}}\sim 0$.
As the majority of the sources on the negative side ($m_{\mathrm{F444W}}-m_{\mathrm{F410M}} < 0$) are distributed at $m_{\mathrm{F444W}}-m_{\mathrm{F410M}} > -0.25$, we adopt 0.25 as the threshold. 
This threshold corresponds to rest-frame EW $\sim$ 660 \AA.
It is noted that this threshold does not necessarily ensure that sources are intrinsically above this selection threshold. This is because EWs are calculated from continuum flux density estimated by spectral energy distribution (SED) fitting and line flux calculated from the F410M flux (see Section~\ref{subsec:emitterproperties}).
Nevertheless, in the following section, we find that most EWs of our final samples are above 660 \AA.

In summary, based on our secure twofold selection, we identify 119 \hb+\oiii\ emitters at $z \sim 7$ (see Figure~\ref{fig:selection}).

\subsection{SED fitting} 
\label{subsec:sedfitting}

The SED was fitted to retrieve $M_{\star}$, dust attenuation ($A_{\mathrm{V}}$), UV slopes ($\beta$), UV magnitudes ($M_{\mathrm{UV}}$), and star-formation rates (SFR) based on {\tt gsf} \citep[ver.1.8.5;][]{Morishita2019}. 
{\tt gsf} is a Python-based SED fitting code that allows us to obtain posterior probabilities for the physical properties of galaxies, such as stellar mass, dust attenuation, metallicity, and the histories of star formation and metallicity enrichment.
Following \cite{Morishita2023}, we generate stellar and interstellar medium templates with ages of 10, 30, 100, 300, 1000 Myrs, dust attenuation of $0 < A_{\mathrm{v}} < 4$, and metallicities of $-2 < \log Z_{\star}/\mathrm{Z_{\odot}} < 0$ (Table~\ref{table:cigale}).
We assume that the stellar and nebular components share the same metallicity. 

{\tt gsf} employs the Markov Chain Monte Carlo ensemble sampler {\tt emcee} \citep{Foreman-Mackey2013} to sample the posterior distributions of the model parameters over $10^{4}$ iterations, using 100 walkers. The uncertainties in the derived physical properties are defined by the 16th -- 84th percentile range of the posterior distributions.

The dust-corrected SFR is derived from \citet{Kennicutt1998} relation and adjusted for the Chabrier IMF by multiplying by a factor of 0.63 \citep[][]{Madau2014}:
\begin{equation}
    \mathrm{SFR_{UV}}\ [\mathrm{M_{\odot}\, yr^{-1}}] = 0.88 \times 10^{-28}\
    \left(\frac{L_{\mathrm{UV, corr}}}{\mathrm{erg\, s^{-1}\, Hz^{-1}}}\right).
\end{equation}
The dust attenuation at 1600 \AA\ is calculated from the UV slope using the relation: $A_{1600} = 4.43 + 1.99\beta_{\mathrm{UV}}$ as in \citet{smit16}.
Dust attenuation low is utilized for the Small Magellanic Cloud (SMC) extinction curve \citep{Gordon2003}.
We also apply the same dust attenuation to the stellar and nebular components.
UV-based SFRs may be overestimated, especially at low metallicities.
However, we do not apply metallicity-dependent corrections, which would complicate the direct comparison between \cite{Morishita2023} and also introduce uncertainties related to the adopted metallicities.

\begin{table*}
\centering
 \caption{Input Parameters for SED Fitting with gsf.}
 \label{table:cigale}
  \begin{tabular}{lcc}\hline
  Parameters & \\\hline\hline
  \quad Initial mass function &  & \cite{Chabrier2003} \\
  \quad Age of the stellar population in the galaxy & & 10, 30, 100, 300, 1000 Myrs \\
  \quad Metallicity &  & $-2 < \log Z_{\star}/\mathrm{Z_{\odot}} < 0$ \\
  \quad Dust attenuation & & $0 < A_{\mathrm{v}} < 4$ \\
  \quad Dust attenuation model &  & SMC\\\hline
  \end{tabular}
\end{table*}

\subsection{\hb\ $+$ \oiii\ line flux and EW}
\label{subsec:emitterproperties}

We derive an \hb+\oiii\ line flux and an EW using the observed F410M magnitude and the continuum flux derived from our best-fit SED.
Since the effective wavelengths of F410M and F444W differ slightly, the $m_{\mathrm{F410M}}-m_{\mathrm{F444W}}$ color is influenced by the intrinsic color of the continuum, potentially affecting our results. 
To mitigate this, we subtract the continuum flux estimated from the SED from the F410M flux to obtain the emission-line flux. 
The stellar and nebular continuum flux ($f_{\mathrm{con}}$) is first derived from the best-fit SED, and the \hb+\oiii\ line flux and EW are then calculated using the following equations:
\begin{equation}
 F_{\mathrm{H\beta+[O_{III}]}} = (f_{\mathrm{F410M}} - f_{\mathrm{con}})\Delta_\mathrm{F410M},
\end{equation}
and
\begin{equation}
 EW_{\mathrm{H\beta+[O_{III}]}} = \frac{F_{\mathrm{H\beta+[O_{III}]}}}{f_{\mathrm{con}}} (1+z)^{-1},
\end{equation}
where $\Delta_\mathrm{F410M}$ is the FWHM of the F410M filter.
To account for the uncertainties of the model, we add the difference between the data and the best-fit SED to the errors of \hb+\oiii\ line flux and EW. 

\subsection{\hb+\oiii\ emission maps} \label{subsec:linedistribution}
The F410M and F444W images allow us to investigate the spatial distribution of the line emission. 
We obtain an emission-line map for individual sources by the following equation
\begin{equation}
    F_{\mathrm{line}}(x_i,y_i) = \frac{f_{\mathrm{F410M}}(x_i,y_i)-f_{\mathrm{F444W}}(x_i,y_i)}{1-\Delta_{\mathrm{F410M}}/\Delta_{\mathrm{F444W}}}\Delta_{\mathrm{F410M}},\label{eq_line}
\end{equation}
and continuum components by
\begin{equation}
    f_{\mathrm{con}} = \frac{f_{\mathrm{F444W}}-f_{\mathrm{F410M}}(\Delta_{\mathrm{F410M}}/\Delta_{\mathrm{F444W}})}{1-\Delta_{\mathrm{F410M}}/\Delta_{\mathrm{F444W}}}\label{eq_con}.
\end{equation}
It should be noted that the F410M image is point-spread function (PSF) matched to the F444W image beforehand.
As mentioned in Section~\ref{subsec:sedfitting}, there is an inherent color in the continuum light between the F410M and F444W filters. 
We do not take the intrinsic color into account here.
However, it does not prominently affect the differences between line and continuum distributions, as F410M is located within the F444W wavelength range, and the continuum light is almost flat within this range (e.g., see Figure~\ref{fig:fig}).

\subsection{Identification of clumpy structures} \label{subsec:identificationclump}

In this study, we utilize additional source extractions in the F150W images to identify clumpy structures. 
These clump structures are commonly identified in rest-frame UV images and optical \citep[e.g.,][]{Elmegreen2005,Elmegreen2007,Ravindranath2006,Genzel2011,Schreiber2011}. The F150W filter corresponds to the rest-frame UV ($\sim$1800\AA) at $z\sim7$.

We carry out source extractions in the F150W image with {\tt Sextractor}, but with a configuration that differs from our source identification in Sec.~\ref{subsec:selection}. 
For our clump identification, we perform source extractions using parameters shown in Table~\ref{table:Sextractor2}.

We define galaxies as clumpy when two or more components are detected within $r<0^{\prime\prime}$.35 from the photocenter in the F150W image, as defined in the original source identification in Section~\ref{subsec:selection}  (see Appendix; Figure~\ref{fig:clumpy}).
However, we removed any contamination and observed clumpy structures within the F090W detection because such a clump at $z\sim7$ should not be visible in the F090W image (see Section~\ref{subsec:selection}).

\begin{table}
\centering
 \caption{Sextractor parameters for identification of clumpy structures.}
 \label{table:Sextractor2}
  \begin{tabular}{lcc}\hline
  Parameters & \\\hline\hline
  \quad DETECT\_MINAREA  &\quad&  10 \\
  \quad DETECT\_THRESH   &\quad&  1.2\\
  \quad ANALYSIS\_THRESH &\quad&  1.2\\
  \quad DEBLEN\_MINCONT  &\quad&  0.0001 \\
  \quad BACK\_SIZE       &\quad&  64\\
  \quad BACK\_FILTERSIZE &\quad&  3\\
  \quad BACKPHOTO\_TYPE  &\quad&  LOCAL\\
  \quad BACKPHOTO\_THICK &\quad&  24\\
  \quad convolution kernel&\quad&gauss\_2.5\_5$\times$5.conv\\\hline
  \end{tabular}
\end{table}

\subsection{Size measurement} 
\label{subsec:sizemeasurement}

For size measurement, we quantify the size of the emitters by measuring the effective radius ($r_{\mathrm{eff}}$) with {\tt Petrofit} \citep{Geda2022}.
We fit the S{\'e}rsic profiles \citep{Sersic1963} to the PSF convolved images of F150W (corresponding to rest-frame UV) and F356W (representing stellar components), and the continuum-subtracted emission-line distributions (\hb+\oiii) for single-component emitters.
We exclude clumpy galaxies defined in Section~\ref{subsec:identificationclump}. 

The PSF images are obtained using {\tt webbpsf} \citep{Perrin2014}.
However, the default output from {\tt webbpsf} shows a narrower PSF profile generated by stars.
To rectify this issue, \cite{Morishita2023} have adjusted the jitter sigma parameter in {\tt webbpsf}, aiming to find the optimal PSF that accurately represents the actual PSF size.
Consequently, they set $0^{\prime\prime}.022$ for the red-channel filters and $0^{\prime\prime}.034$ for the blue-channel filters. 
This adjustment significantly improved the accuracy of their work. In this study, we adopt the same jitter sigma parameter.

{\tt Petrofit} requires initial values for its fitting parameters. 
To estimate realistic errors in the output parameters, we repeat the fit 10 times for each galaxy, under different initial conditions.
We add random fluctuations of 1$\sigma$ to the photometric center and the magnitude listed in our photometric catalog (Section~\ref{subsec:selection}). 
For the S{\'e}rcic index, since we do not have any a priori estimate, we assign random values ranging from 0.1 to 4 in each iteration.
The errors in the output parameters are then set to the standard deviation of the ten measurements. 
We exclude sources with low signal-to-noise ratio (S/N) $<$ 3 from our size analysis in this study, as we find that the measurement uncertainties are too large. 
We obtain 81 measurements for the \hb+\oiii\ emission component, 27 for the stellar, and 56 for the rest-frame UV. 


\begin{figure}
\begin{center} 
\includegraphics[width=82mm]{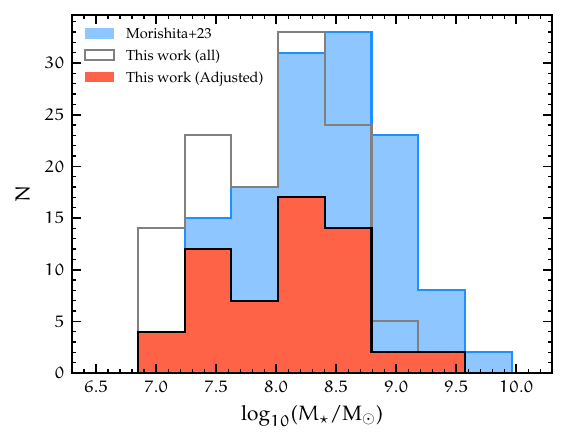}
\end{center} 
\caption{
Adjusted stellar mass distributions of our EELGs (red) and UV-selected galaxies  \citep[blue;][]{Morishita2023}.
The comparison is made by showing only the UV- UV-selected galaxies that are selected in the same fields as our emitters. The gray histogram shows the mass distribution of all our emitters. The EELGs show a skewed distribution toward the low-mass end ($p=0.02$).
} 
\label{fig:hist_Ms}
\end{figure}

\begin{figure*}
\begin{minipage}{0.5\hsize} 
\begin{center} 
\includegraphics[width=82mm]{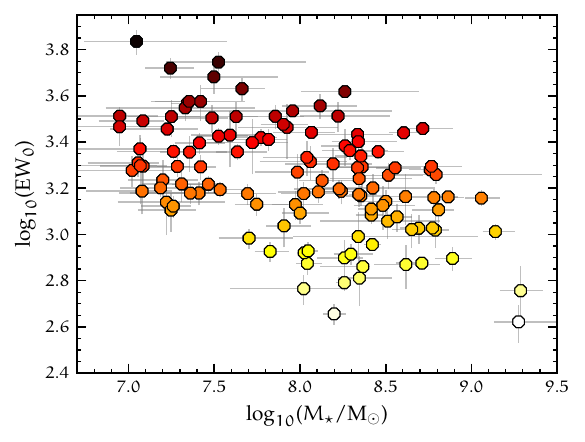}
\end{center} 
\end{minipage} 
\begin{minipage}{0.5\hsize} 
\begin{center} 
\includegraphics[width=82mm]{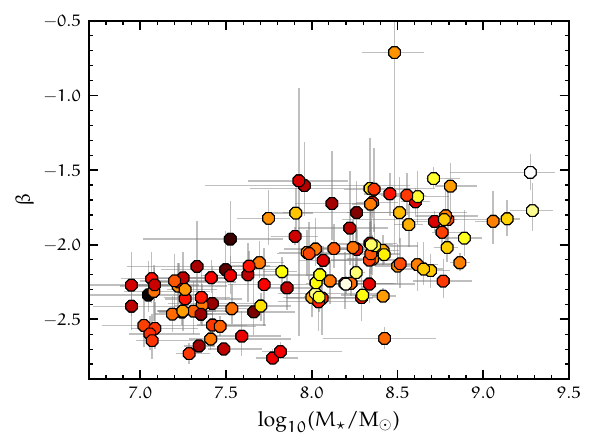}
\end{center} 
\end{minipage} 
\begin{minipage}{0.5\hsize} 
\begin{center} 
\includegraphics[width=82mm]{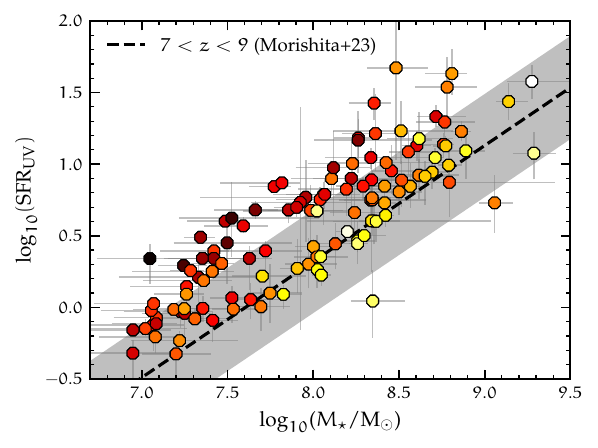}
\end{center} 
\end{minipage} 
\begin{minipage}{0.5\hsize} 
\begin{center} 
\includegraphics[width=82mm]{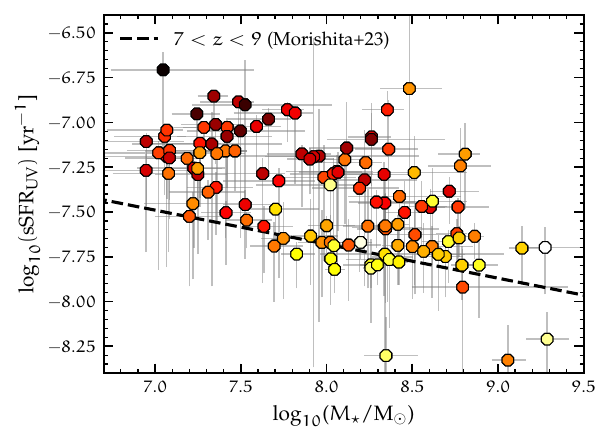}
\end{center} 
\end{minipage} 
\begin{minipage}{0.5\hsize} 
\begin{center} 
\includegraphics[width=82mm]{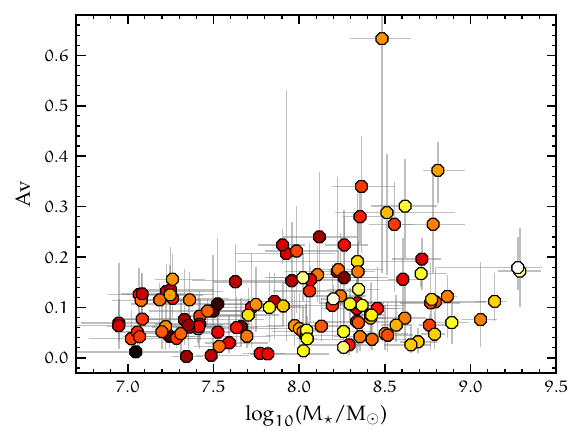}
\end{center} 
\end{minipage} 
\begin{minipage}{0.5\hsize} 
\begin{center} 
\includegraphics[width=82mm]{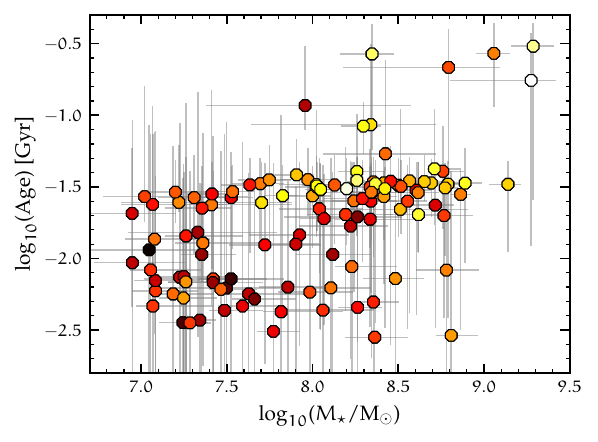}
\end{center} 
\end{minipage} 
\caption{Physical properties of \hb+\oiii\ emitters at z $\sim$ 7 plotted against their stellar mass ($M_{\star}$). 
(Top left) The rest-frame EW$_0$ -- $M_{\star}$ relation. The color of each point corresponds to the rest-frame EW.
(Top right) The $beta$ -- $M_{\star}$ relation.
(Middle left) The SFR -- $M_{\star}$ relation. The dotted line is the relationship derived for Lyman break galaxies at $7.2 < z < 9.7$ \citep{Morishita2023}. The shaded region shows the intrinsic scatter (0.36 dex). 
(Middle right) The sSFR -- $M_{\star}$ relation.
(Bottom left) The dust attenuation ($A_{\mathrm{V}}$) -- $M_{\star}$ relation.
(Bottom right) The age -- $M_{\star}$ relation.
It is clear that the majority of EELG sources which are about 50\% of the Lyman beak galaxy population in these fields, have high specific star-formation rates, low dust content, and young stellar ages. 
}
\label{fig:Ms} 
\end{figure*}

\section{Results} \label{sec:results}
\subsection{Properties of \hb\ $+$ \oiii\ emitters at $z$ $\sim$ 7} \label{subsec:highEWemitters}

We investigate the properties of the 119 \hb\ $+$ \oiii\ emitters at $z$ $\sim$ 7 identified in Section~\ref{subsec:selection}. 
Our samples span the stellar mass range of $\sim10^7$--$10^
{9}\,\mathrm{M_\odot}$ (Figure~\ref{fig:hist_Ms}). 
The distribution is skewed towards lower mass compared to UV-selected sources at similar redshifts \citep[][]{Morishita2023}. A Kolmogorov-Smirnov (K-S) test raises the $p$-value of $0.02$.

Figure~\ref{fig:Ms} illustrates the selected sources in various parameter spaces, namely stellar mass ($M_{\star}$), EW$_0$, Av, and $\beta$. 
A significant portion of our samples are situated on the upper side of the main sequence (MS) indicated by \cite{Morishita2024}. 
This trend is likely attributed to our sample selection, which used the color excess in F410M $-$ F444W, as we tend to detect extreme emitters above the MS more readily.

The selected sources exhibit a wide range in \hb\ $+$ \oiii\ rest-frame EW$_0$, specifically 420 $<$ EW$_{0}$ / \AA\ $<$ 7200 (Figure~\ref{fig:Ms}, Top left).
An average EW of our emitters is EW$_{0}$ = 1744 \AA.
Of note, we discovered 19 galaxies with very high EWs,  $>$ 3000 \AA\ (high-EW emitters; see Figure~\ref{fig:ImageHEWsSEDs_s} -- \ref{fig:ImageHEWsSEDs_e}).
Similar high-EW emitters have surfaced in several studies \citep[e.g.,][]{DeBarros2019,Endsley2023a,Endsley2023b,Matthee2023}. 
Our high-EW emitters are characterized by their low masses ($\log M_{\star} / M_\odot < \sim 10^{7.5}$), and low dust contents ($A_{\mathrm{v}}\sim 0.1$ mag), and blue colors ($\beta \sim -2.2$), as shown in Figure~\ref{fig:Ms}.

As mentioned in Section~\ref{sec:intro}, we define high-EW galaxies as those with a rest-frame \hb\ + \oiii\ EW greater than 3000 \AA, and EELGs as galaxies with a rest-frame \hb\ + \oiii\ EW greater than 1000 \AA.
The fractions of EELGs and high-EW emitters, based on UV-selected (Lyman-break) galaxies, are 32.8\% and 3.7\% respectively.
Hence, high-EW emitters are rare entities at $z\sim7$.
Almost all emitters with $\mathrm{EW_0}>3000$ \AA\ are located above the star-forming MS (18/19 objects) \citep[$>$0.37 dex][]{Morishita2024}.
This observation suggests that high-EW emitters are potentially engaged in a burst phase of star-forming activities. 

On the other hand, we observe a fraction of our emitters on the MS derived for more general UV-selected SFGs (49\% = 58/119). 
This implies that, although our selection involves an additional step using the F410M color excess, some galaxies (with relatively small EWs) overlap with the UV-selected galaxy samples.

\subsection{Size Distributions of the \hb+\oiii\ emitters}\label{size}

\begin{figure}
\begin{center} 
\includegraphics[width=82mm]{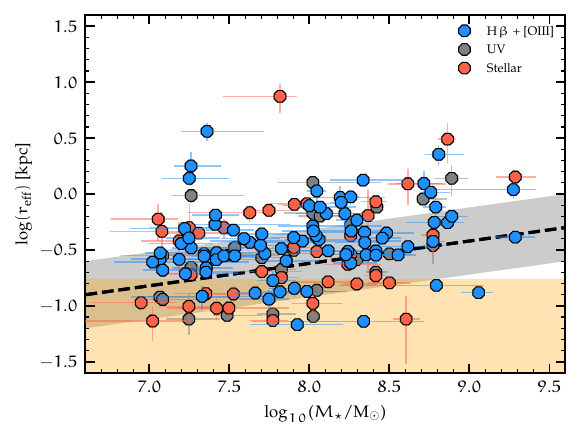}
\end{center} 
\caption{
High-EW emitters from our sample on the mass--size relation diagram (effective radius ($r_{\mathrm{eff}}$) vs. $M_{\star}$). Sizes measured in different wavelengths are shown by different colors (i.e.\ \hb+\oiii\ line in blue, rest-frame UV in gray, and stellar mass component in red). The dotted line and shaded region show the mass--size relation and its scatter derived for Lyman-break galaxies at $7.2 < z < 9.7$ \citep{Morishita2023}. The shaded orange region shows the physical size of the point spread function FWHM/2 for NIRCam, translated into the physical scale at $z = 7.3$.
}
\label{fig:Masssize}
\end{figure}

Figure~\ref{fig:Masssize} shows the size distribution of the \hb+\oiii\ emitters.
We examine the effective (half-light) radii of the  \hb+\oiii emission line, stellar continuum, and UV(F150W) light.
The sizes of the stellar components are measured in the \hb+\oiii-subtracted image, which is derived by following Eq.~\ref{eq_con}.
We use the F150W flux as the UV component and measure their sizes. The pivot wavelength of the F150W filter corresponds to about rest-frame 1810 \AA\ at $z=7.3$. 

The \hb+\oiii\ sizes of our emitters are consistent with the rest-frame UV sizes of the UV-selected galaxies \citep{Morishita2023}.
Some emitters have \hb+\oiii\ and stellar sizes that are more than 0.5 dex greater than this relationship.
We should note that Figure~\ref{fig:Masssize} is presented only to include reliable results (size measurement above $>3\sigma$) for each component.
As a result, Figure~\ref{fig:Masssize} might not display extended galaxies, weak emission-line galaxies, low-mass galaxies, or dusty galaxies. 
Extended galaxies and weak emission-line galaxies could have low S/N, rendering them undetectable.  
Low-mass galaxies might not be visible due to their faint continuum, and dusty galaxies could be too dim because of dust attenuation.

\subsection{Spatial distribution of \hb\ $+$ \oiii\ line emission}
\label{subsec:linemap}

To investigate the origin of high EW, we compare the  \hb\ $+$ \oiii\ emission-line distribution with F150W (corresponding to the rest-frame UV) and F356W (corresponding roughly to the stellar mass) images.
We initially ascertain the peak position using {\tt photutils.detection.find\_peaks} \citep{Photutils} for all three components.

Next, we determine the central position of these objects utilizing {\tt photutils.detection.DAOStarFinder}, measuring peak position values from the local density, is characterized by a peak amplitude and a size and shape similar to the defined 2D Gaussian kernel. 
The source-finding step is skipped since we are aware of the approximate position with the found peak. 
However, we defined the error by varying the initial value by a normal distribution with half the average size of the emitter (1.245 pixels) as the standard deviation, repeated 100 times.

\begin{figure*}
\begin{minipage}{0.5\hsize} 
\begin{center} 
\includegraphics[width=90mm]{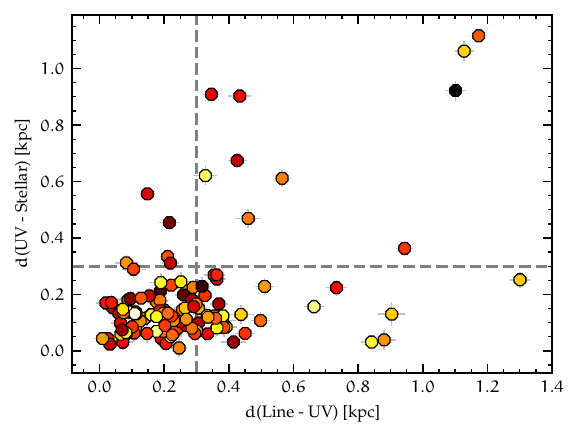}
\end{center} 
\end{minipage}
\begin{minipage}{0.33\hsize} 
\begin{center} 
\includegraphics[width=90mm]{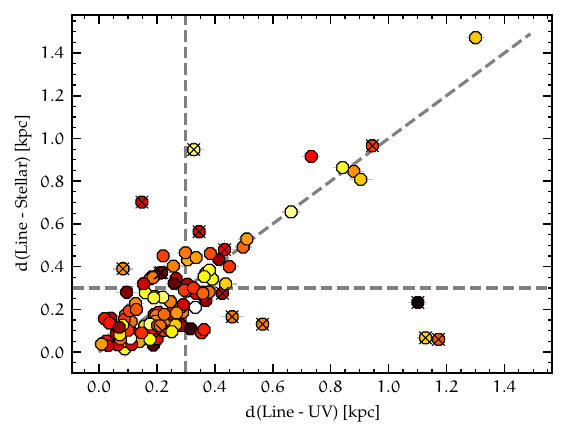}
\end{center} 
\end{minipage}
\caption{(Left) The relationship between the centroid distances of UV and stellar mass component ($d(\mathrm{UV-stellar})$) and the distances of \hb + \oiii\ line and UV component (($d(\mathrm{line-UV})$)). Dashed lines show the criteria of galaxies with offset (0.3 kpc, see Figure~\ref{fig:offset} - \ref{fig:offset3}). (Right) The relationship between the centroid distances of \hb + \oiii line and stellar component ($d(\mathrm{line-stellar})$) and ($d(\mathrm{line-UV})$). The diagonal dashed line indicates a one-to-one ratio. The cross marks are galaxies with $d(\mathrm{UV-stellar})<0.3$ kpc, mainly galaxies with a clump or poor SNR in images. 13 sources show large offsets of the line emission from the UV and stellar mass component suggesting the presence of ionized outflows and/or shocks.}
\label{fig:distance} 
\end{figure*}

Figure~\ref{fig:distance} illustrates the distance from the stellar component as a function of the distance between the emission line and the UV and stellar components.
Clumpy galaxies may potentially have unusual values due to variances in dust attenuation, star formation, and stellar distribution between each clump.
Emitters with $d(\mathrm{UV-stellar}>0.3)$ kpc are likely to be imaged with a poor S/N or clumpy galaxies. 

Excluding these objects, $d(\mathrm{Line-stellar})$ and $d(\mathrm{Line-UV}$ appear to be correlated,  despite having scatter (see Figure~\ref{fig:distance} right).
The \hb\ + \oiii\ line distribution with offset might be due to the ionized gas flows caused by strong feedback from massive stars and/or AGNs.
Given that our targets are mostly low-mass systems  (Figure~\ref{fig:Ms}), it seems likely that they are more influenced by such feedback processes.
We find a clear offset in 13 objects with $d\mathrm{(Line-UV)} > 0.3\ \mathrm{kpc}$ and $d\mathrm{(Line-UV)} > 0.3\ \mathrm{kpc}$, but  $d\mathrm{(UV-stellar)} < 0.3\ \mathrm{kpc}$.
Physical proprieties of \hb\ + \oiii\ emitter with spatial offset are shown in Table~\ref{tab:offset}.
The origins of this offset are discussed in Section~\ref{sec:discussion}.

\subsection{Clumpy structure} 
\label{subsec:identificationclump}

Clumpy structures are observed in 38 $\pm$ 8\% of our samples. 
To compare the nature of clumpy galaxies among our EELGs with UV-selected galaxies at similar redshifts, we also make a sample of clumpy galaxies for the latter, finding clumpy structures in 25 $\pm$ 3\% of the UV-selected galaxies.
However, it is noteworthy that our EELGs appear to have higher clumpy fractions than the normal galaxies.
One possible explanation is that their high EWs are partially stimulated by the violent gravitational instability of the disk or mergers related to the clumpy galaxies \citep[e.g.,][]{Dekel2009,Dekel2013,DiMatteo2008,Guo2015,Shibuya2016,Tadaki2018}.

\section{Discussion: Origins of extreme \hb\ $+$ \oiii\ emission lines}
\label{sec:discussion}

We have observed galaxies with notably high EWs, sometimes as high as $>$ 3000 \AA.
Physical proprieties and SEDs of \hb\ + \oiii\ emitter with rest-frame \hb\ + \oiii\ EW $>$ 3000 \AA are shown in Table~\ref{tab:HEW} and Figure~\ref{fig:ImageHEWsSEDs_s} -- \ref{fig:ImageHEWsSEDs_e}.
High-EW emitters have been documented in the literature \citep{Labbe2013, Smit2014,Roberts-Borsani2016,DeBarros2019,Morishita20} and in recent JWST spectroscopy \citep[e.g.,][]{Gupta2023,Matthee2023}. 

The origins of extreme nebular emissions include: (1) AGN, (2) Stellar photoionization in the vicinity of \hii\ regions, and (3) Radiative shocks powered by outflows either from AGN or massive stars, as schematized in Figure~\ref{fig:fig}. 
With the imaging data set used here, it is challenging to understand the exact origin of all individual sources. 
On the other hand, we have found 13 emitters showing offset \hb\ $+$ \oiii\ emission compared to the stellar and UV components (Section~\ref{subsec:identificationclump}; Figure~\ref{fig:distance}).
This can be attributed unequivocally to the third scenario because these three components would align when ionization is primarily driven by star formation or AGN.

\subsection{Bursty star-forming activities in Extreme Emission-line Galaxies}
\label{subsec:dis_star-burst}

\begin{figure*}
\begin{minipage}{0.5\hsize} 
\begin{center} 
\includegraphics[width=90mm]{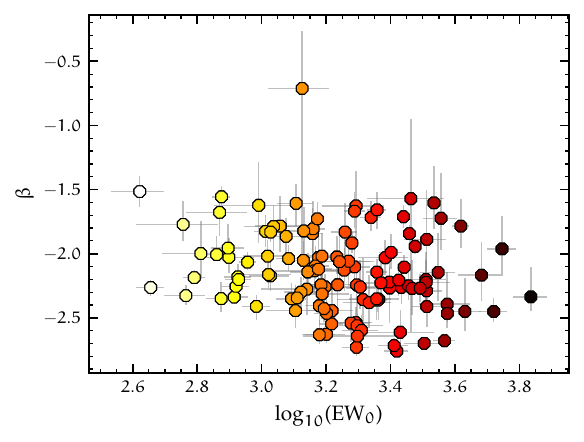}
\end{center} 
\end{minipage} 
\begin{minipage}{0.5\hsize} 
\begin{center} 
\includegraphics[width=75mm]{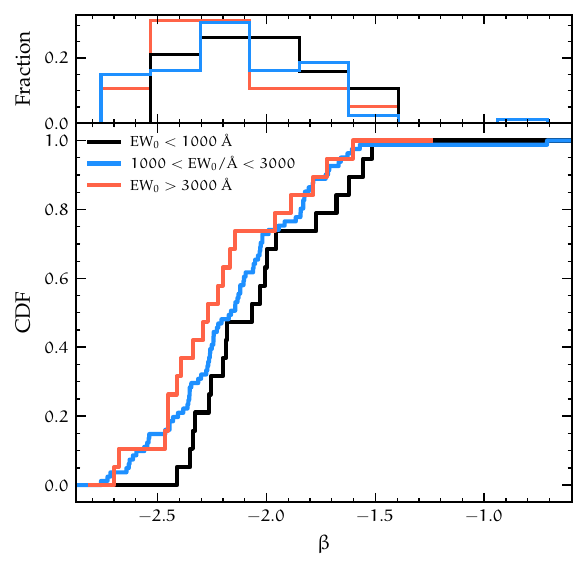}
\end{center} 
\end{minipage}
\caption{(Left) Sample distributions on the $\beta$--rest-frame EW$_0$ plane. $\beta$ is only mildly dependent on EW$_0$. (Right) Histogram of $\beta$. The samples are divided into three groups: those with EW$_0$ values above 3000 \AA, those below 1000 \AA, and an intermediate group. A K-S test comparing the samples above 3000 \AA\ and below 1000 \AA\ revealed no statistically significant difference.}
\label{fig:beta} 
\end{figure*}

Approximately 50 \% of our samples are above the star formation main sequence (SFMS; $>0.3$ dex). 
Our EELG sample is 84\% (=16/19) located on the upper side of the MS ($\Delta \mathrm{MS}>0.3$).
We also found that a significant portion (44\%) of our samples display clumpy features (see Section~\ref{subsec:identificationclump}).
These clumpy structures are presumed to be formed by the fragmentation of gas clouds in cases of gravitational instability of turbulent disks and clumpy streams \citep[e.g.,][]{noguchi98,noguchi99,Dekel2009,Dekel2013} and/or gas-rich mergers \citep[e.g.,][]{DiMatteo2008}.

In this study,  the difference between the photometry-derived EW and the model value is included as part of the measurement uncertainty (see Section~\ref{subsec:sedfitting}).
In other words, this indicates that the synthesis model can reproduce the high EW observed. 
Our investigation of the fitting results shows that EELG candidates with EWs exceeding 3000 \AA\ correspond to low-mass, young galaxies undergoing starburst activity, as illustrated in Figure~\ref{fig:Ms}.
In particular, the age parameter is markedly different from that of other populations. 
This suggests that low-mass ($M_{\star}\sim10^{7.3}$ $\mathrm{M_{\odot}}$), youth, and an active starburst phase may be key factors in achieving such high EWs.

In \cite{Saxena2024}, the UV slope $\beta$ was measured from synthetic spectra as a function of time after a burst, assuming a Chabrier IMF \citep[see Figure 8 in][]{Saxena2024}. The results show that for metallicities in the range $-2.0 < \log(Z/\mathrm{Z_{\odot}}) < -0.5$, $\beta$ is distributed roughly around $-2.5$. As expected, since $\beta$ is affected by dust attenuation, the values tend to be higher in more massive galaxies with EW below 3000,\AA. These are broadly consistent with the observed $\beta$ distribution of our samples \citep[see Figure 2 in][]{Saxena2024}.

Interestingly, however, no strong dependence of $\beta$ on EW was found; there is no statistically significant difference between galaxies with EW below 1000,\AA\ and those with EW above 3000,\AA, as tested using the K–S test (Figure~\ref{fig:beta}). One possible explanation for this behavior is either the rapid buildup of dust reservoirs or a substantial contribution from nebular continuum emission to the rest-frame UV slope. Especially, \cite{Saxena2024} demonstrated that nebular continuum emission would redden the UV spectrum when the gas temperature exceeds $\sim15000$\,K. Therefore, the observed $\beta$ distribution of our EELGs is not inconsistent with the UV continuum being dominated by the nebular emission of high temperature.

The resolution of NIRCam has shown that distinguishing clumpy structures from galaxy mergers at $z\sim7$ remains a significant challenge due to the small apparent sizes of galaxies at these redshifts. Previous studies have suggested that a fraction of EELGs are either undergoing mergers or exhibit clear signs of interaction \citep{Gupta2023}. 
Like clumps, mergers, and galaxy interactions can significantly reduce a galaxy’s angular momentum, funneling gas toward the center and triggering starburst events. 
However, in our EELG sample, only three galaxies show strong emission located between clumps, suggesting that the fraction of ongoing mergers is small. 
High EWs are not exhibited in any of our samples.

High-EW emitters are proposed to produce substantial LyC photons in association with active star formation, contributing to cosmic reionization.
For example, \cite{Saxena2023} reported a strong \lya\ emitter at $z=7.28$ with JWST G140M grating observation (JADES-GS+53.016746-27.7720).
The emitter has EW(\hb+\oiii) $\sim$ 1300 \AA\ and EW(\lya) = 388.0 $\pm$ 88.8 \AA.
This is identified as one of the high-EW \lya\ emitters during the epoch of reionization.
Notably, this object also exhibits a high \lya\ escape fraction ($>$70\%), low metallicity, and high ionization state -- characteristics determined by line diagnostics.
The Green Pea galaxies, bearing similarity to our samples, also exhibit high \lya\ escape fractions \citep[e.g.,][]{Henry2015,Yang2017,Izotov2020,Chisholm2022,Amorin2024,Jung2024}.
Moreover, \cite{Kim2023} revealed that galaxies, characterized by spatially resolved LyC-leaking regions,  tend to portray a prominent blue UV slope, an elevated ionization state, strong oxygen emission, and a high  \lya\ escape fraction. 
Our high-EW emitters are also highly likely to possess such a LyC-leaking region, due to a comparably dynamic star formation. 
However, spatially resolving these sources proves challenging within the redshift range of our interest and necessitates the assistance of robust gravitational lensing \citep[e.g.,][]{vanzella23}.

\subsection{Potential contribution from Active Galactic Nucleus}
\label{subsec:dis_emitters}

While our dataset does not allow us to directly investigate the contribution of AGN, we aim to estimate an upper limit of the AGN contribution to the observed EW under a few reasonable assumptions. 

As a simple test, we assume that each emitter has a faint AGN with $\log_{10}$(\oiii/\hb) $=$ 0.5 and $M_{\mathrm{BH}}/M_{\star}=0.01$.
The latter value is determined based on the faint AGN at $6<z<9$ reported in \cite{Harikane2023}, and the emission-line ratio is taken as the approximate minimum value for faint AGN.

Based on these assumptions, we estimate the black hole mass ($M_{\mathrm{BH}}$) using the stellar mass from our SED analysis. 
We then calculate the Eddington luminosity ($L_{\mathrm{Edd}}$) from the relationship between the $L_{\mathrm{Edd}}$ and $M_{\mathrm{BH}}$; $L_{\mathrm{Edd}}=1.26\times10^{38}\ (M_{\mathrm{BH}}/\mathrm{M_{\odot}})$ erg/s.
Assuming the Eddington ratio to be 1, we can estimate \oiii\ emission luminosity from bolometric luminosity $L_{\mathrm{bol}}$ (= $L_{\mathrm{Edd}}$) = 3500 $\times$ $L_{\mathrm{[OIII]}}$ \citep{Heckman2004}. 

Using the assumption $\log_{10}$(\oiii/\hb) $=$ 0.5, we estimate the \hb+\oiii\ emission-line flux.
Finally, we derive the AGN-driven EW using the same $f_{\mathrm{con}}$ when we calculate the rest-frame \hb+\oiii\ EW.

\begin{figure}
\begin{center} 
\includegraphics[width=82mm]{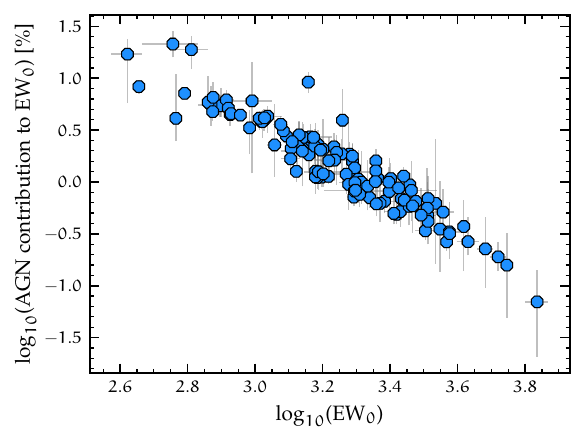}
\end{center} 
\caption{Prediction for the maximum contribution from AGN to observed EW$_{0}$. We assume that all sources host an SMBH with a typical $M_{\mathrm{BH}}/M_{\star}$ from the $M_*$--$M_{\rm BH}$ relation. AGN contribution decreases with EW. In particular, high-EW emitters have little AGN contribution. This suggests that starbursting or shock heating may be the main ionization source for most EELGs (see Sec.~\ref{subsec:dis_emitters}).}
\label{fig:fraction_EW_AGN}
\end{figure}

The estimated potential contribution to EW$_{0}$ by AGN is shown in Figure~\ref{fig:fraction_EW_AGN}.
AGN contribution of EELGs ($EW_{0}>1000$\AA) decreases with the observed EW, but AGN contributions remain less than 10\% in most cases.
Especially, high-EW emitters ($EW_{0}>3000$\AA) are below 0.7\%, suggesting that AGN is likely to have little contribution.
This EW dependence primarily arises from the observed negative correlation with stellar mass (Fig.~\ref{fig:Ms}). This correlation, in combination with the mass-dependent $L_{\mathrm{[OIII]}}$ assumed above, results in the trend of the AGN contribution seen in the figure.

The result suggests that AGN is unlikely to be the dominant ionizing source over the EW range of our sample. 
Instead, other non-AGN sources, such as starbursting or shock heating from massive stars, are likely the candidates, a similar conclusion as reported in, e.g., \citet[][]{zhang24}.
With the current data, however, it is challenging to firmly determine the origin. It is plausible that a complex interplay of these phenomena leads to the observed high EW.
Further investigation of these interesting objects through follow-up observations should place limits on the mechanisms that lead to high EW.

\subsection{Offset Emissions due to shocks from strong outflows}\label{sec:offset}
\label{subsec:ionization-source}

In Section~\ref{subsec:linemap}, we investigate $d\mathrm{(UV - stellar)}$ and $d\mathrm{(Line-stellar)}$ as a function of $d\mathrm{(Line-UV)}$ (see Figure~\ref{fig:distance}).
We identify 13 objects with a clear offset of $d\mathrm{(Line-UV)} > 0.3\ \mathrm{kpc}$ and $d\mathrm{(Line-UV)} > 0.3\ \mathrm{kpc}$, but  $d\mathrm{(UV-stellar)} < 0.3\ \mathrm{kpc}$ (see also Figure~\ref{fig:offset} -- \ref{fig:offset3}).
Galaxies with $d\mathrm{(UV-stellar)} > 0.3\ \mathrm{kpc}$ are mainly clumpy galaxies (see Section~\ref{subsec:linemap}).

These 13 objects with clear offsets are likely candidates of being heated by shocks originating from massive stars or AGN, but unlikely the latter, as discussed in Section~\ref{subsec:dis_emitters}.
If 13 objects are affected by the photoionization from AGN or stellar photoionization in the vicinity of the \hii\ regions shown in Figure~\ref{fig:fig}, offsets between UV, stellar continuum, and line distribution are not expected.
However, we should note that the offset does not rule out the possibility of photoionization from AGN or \hii\ regions.
Given that spatial offset along the line-of-sight direction cannot be distinguished, we estimate an upper limit that at least 11\% of our EELGs are significantly driven by shock heating mechanisms.

Our targets have a high specific SFR. 
Thus, they are expected to have strong outflows.
The nature of these 13 objects is no more bursty than the other emitters (see Table~\ref{tab:offset}).
This result may be due to UV-based SFRs that trace the long timescale of star formation.
Therefore, these objects may not be reflected in the SFR, although they may be undergoing bursts of star formation.

To investigate whether this offset is of shock origin, the kinematics of the gas should be examined by additional integral field units (IFU) observations.

\section{Overdensity of Extreme Emission-line Galaxies}
\label{sec:dis_overdensity}

\begin{figure*}
\begin{minipage}{0.5\hsize} 
\begin{center} 
\includegraphics[width=82.5mm]{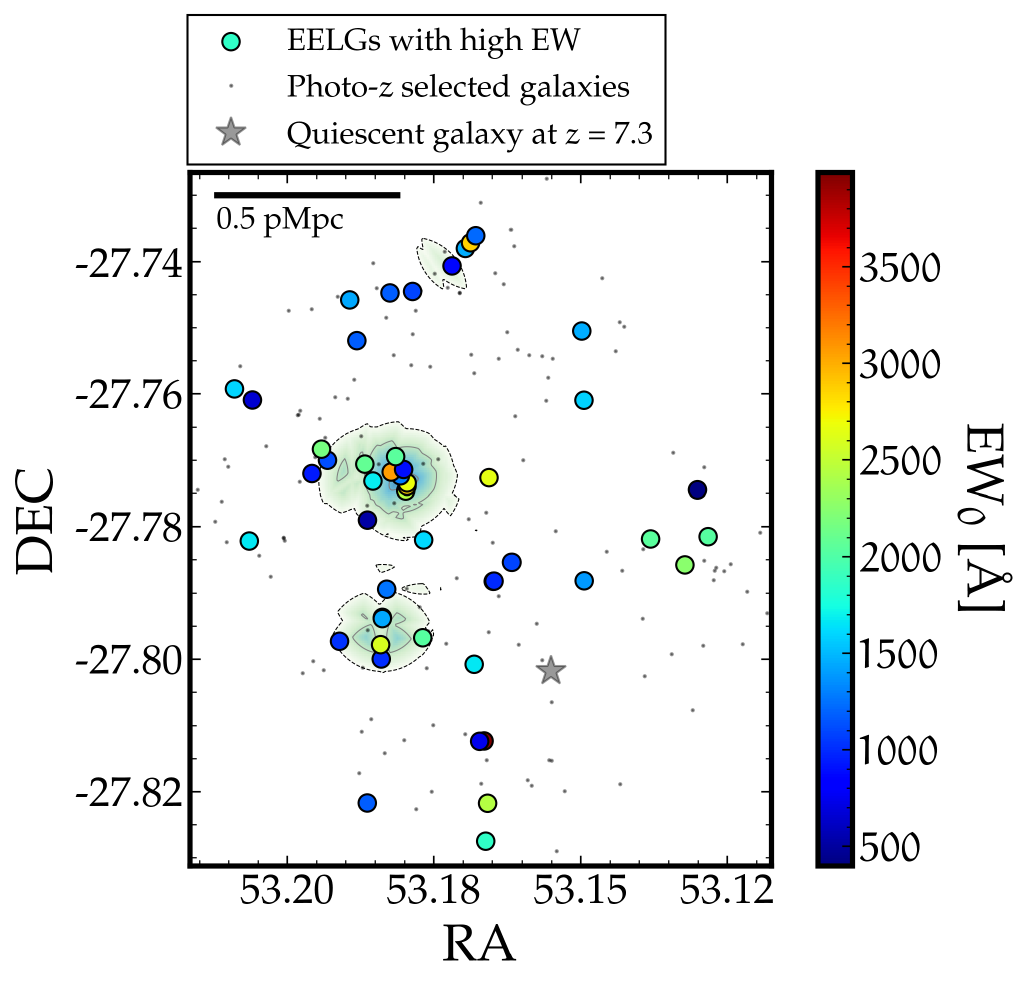}
\end{center} 
\end{minipage} 
\begin{minipage}{0.5\hsize} 
\begin{center} 
\includegraphics[width=82.5mm]{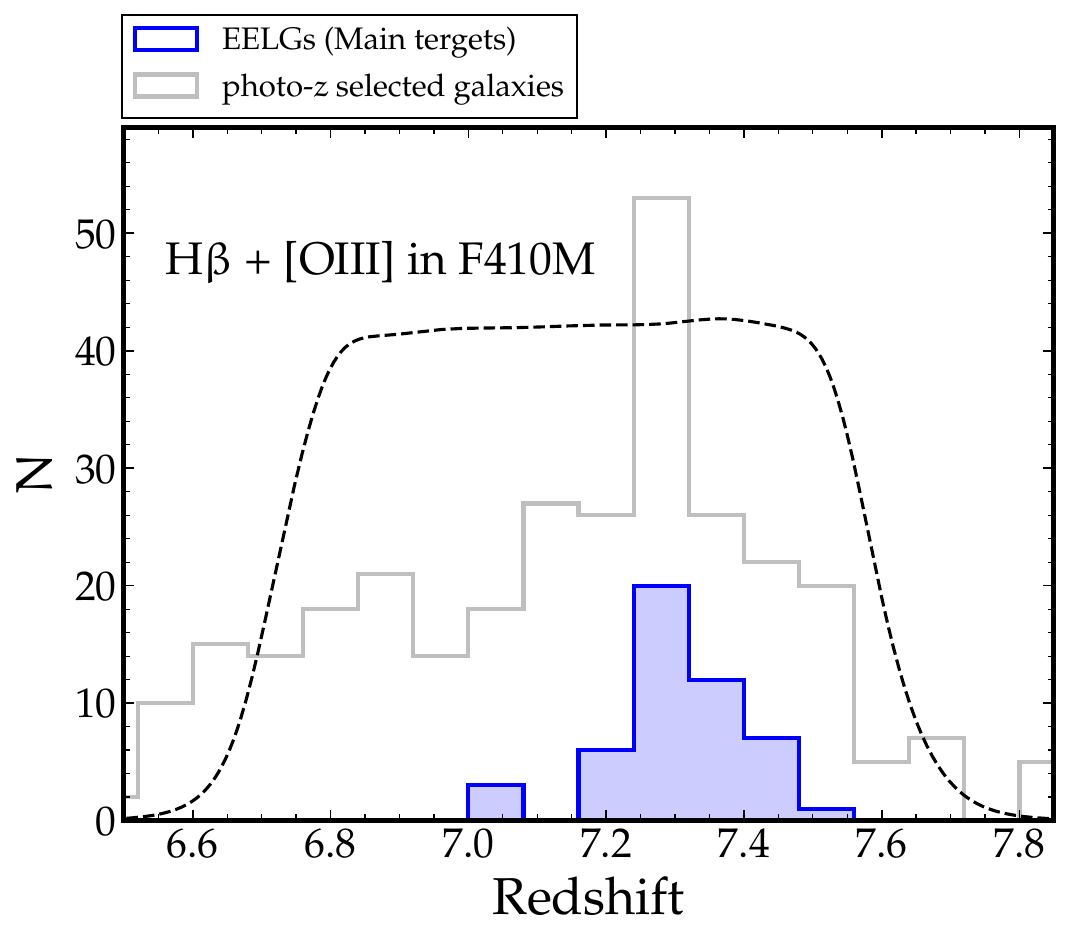}
\end{center} 
\end{minipage}
\caption{(Left) Spatial distribution of \hb+\oiii\ emitters with the color showing the EW values. 
The contours show the mean projected distance $\bar{b}_{\mathrm{5th}}$ of 150, 100, and 50 kpc in the outside-in order, defined by $2 \sqrt{1/(\pi \Sigma_{\mathrm{5th}})}$, where $\Sigma_{\mathrm{5th}}$ is given as $4/\pi r_{\mathrm{5th}}^2$. This indicates the local environment based on the 2D number density of galaxies, and our emitters are found in extremely dense regions. (Right) Photometric redshift distribution of JADES sources. It has a sharp peak at $z\sim7.3$. The redshift probability at $z > 6$ is $>$ 80 \% for all of our sample galaxies, securing low contamination by low-$z$ galaxies. The response curve of the medium-band filter F410M, which captures the flux excess by \hb+\oiii\ emission line, is shown by the dashed line.}
\label{fig:overdensity} 
\end{figure*}

Among our EELG samples collected from various fields, we discovered an overdensity of EELGs in the JADES GOODS-South field. 
The spatial distribution and photometric redshift distribution of our emitters are illustrated in Figure~\ref{fig:overdensity}.
We calculated the overdensity index, $\delta_{\mathrm{gal}}$, to quantify the degree of overdensity, which is defined as follows:
\begin{equation}
    \delta_{\mathrm{gal}} = \frac{n_{\mathrm{gal}} - \ev{ n_{\mathrm{gal}}}}{\ev{n_{\mathrm{gal}}}},
\end{equation}
where $n_{\mathrm{gal}}$ is the number of galaxies within a 240 physical kpc diameter aperture and $\ev{n_{\mathrm{gal}}}$ is the average value of $n_{\mathrm{gal}}$.
Figure~\ref{fig:overdensity} demonstrates overdensity regions where the mean projected distance is less than 100 kpc within approximately 120 kpc.
We count the number of galaxies with $m_{\mathrm{F410M}}$ $>$ 27.5 magnitudes within the aperture by setting a 240 physical kpc diameter aperture.
To explore the spatial clustering of high-EW emitters, we set a threshold of $\mathrm{EW_{0}}$ $>$ 1000 \AA.
The dense regions exhibit an overdensity, with a $\delta_{\mathrm{gal}}$ $\sim$ 50 of EELGs.
\footnote{Using the 1 pMpc diameter aperture, EELG overdensity $\delta_{\mathrm{gal}}$ $\sim$ 10.}
We again emphasize that our sample is secure, as the integrated probability of redshift at $z$ $>$ 6 is $>$ 80\% for all the sample galaxies.
Also, the average value is less than the standard deviation in all regions.

In \cite{Helton2023}, the same overdense region was reported at $z=7.265\pm0.007$, with $\delta_{\mathrm{gal}} = 6.631 \pm 1.322$. 
They estimated halo mass at approximately $\log_{10}(M_{\mathrm{Halo}}/\mathrm{M_{\odot}})$ = 11.683 $\pm$ 0.106, based on the stellar to halo mass relation derived from the semiempirical UniverseMachine model \citep{Behroozi2019}. 
Thus, the spectroscopically confirmed overdense region contains more galaxies with high EW than in the field.

In addition, JADES-GS-z7-01-QU, a low-mass quiescent galaxy at $z$ = 7.3 \citep[][]{Looser2023}, is located $\sim0.5$\,pMpc from the overdensity region (Figure~\ref{fig:overdensity}). 
One may speculate that the observed quiescence of JADES-GS-z7-01-QU is related to the proximity of the overdensity, where intense SFR may produce radiation-driven dusty outflows \citep[e.g.,][]{Gelli2023}.

As seen in these two examples, the clustering of EELGs could be a signpost of interesting regions that host large ionization bubbles and quiescent populations.

\section{Conclusion}
\label{sec:conclusion}

In this study, we identified 119 \hb+\oiii\ emitters with EW$_{0}$ $>$ 420\AA\, at $z \sim 7$ in the public fields of several JWST Cycle 1 programs.
Those emitters were chosen based on the flux excess in the F410M medium-band filter that would capture \hb+\oiii\ emission lines for the redshift range, compared to the overlapping broadband filter F444W, which samples the underlying continuum.  
We conducted a systematic survey of EELGs ($>1000$ \AA) and examined their environmental dependence. 
The key findings are summarized below:

\begin{enumerate}
\item 
The emission-line galaxies have stellar masses of 10$^{6.9}$ $<$ $M_{\star}/\mathrm{M_{\odot}}$ $<$ 10$^{9.3}$, and emission-line equivalent widths of 420 $<$ EW$_{0}$ / \AA\ $<$ 6850 (Figure~\ref{fig:Ms}).
EELGs ($\mathrm{EW_{0}}>1000$ \AA) are low-mass ($\sim10^{7.9}\,\mathrm{M_{\odot}}$), young (characterized by a blue color with $\beta\sim-2.2$), and low dust content ($A_{\mathrm{V}}\sim0.1$) populations.
Similarly, high-EW emitters ($\mathrm{EW_{0}}>3000$ \AA) are also low-mass ($\sim10^{7.5}\,\mathrm{M_{\odot}}$), young ( $\beta\sim-2.2$), and low dust content ($A_{\mathrm{V}}\sim0.1$) populations. 
Moreover, most high-EW emitters (18/19 objects) are in the starbursting phase, elevated for $\simgt0.3$\,dex above the star-forming main sequence.
\\

\item Due to the efficiency of our line-mapping technique, this work provides a first statistical investigation of emission-line spatial distribution and their sizes at $z$=7.
In some of them, we found a spatial offset between the \hb+\oiii\ line emissions and stellar mass and/or rest-frame UV components.
Our targets, especially high-EW emitters, have high specific SFRs and are thereby expected to have strong outflows.
We find 13 objects with \hb\ + \oiii\ line distributions offset from the stellar and UV components ($>$ 0.3 kpc), which may have been affected by shock heating from massive stars and/or AGN.
The diverse differences in those light/mass distributions observed in our samples point to different underlying physical processes. 
Further investigation is warranted to unravel the specific mechanisms driving these differences and their implications for galaxy evolution.\\

\item  We argued that high-EW emitters may originate from starbursts. 
The synthetic model used for our SED modeling could reproduce a high EW at least for objects with low-mass, young galaxies undergoing starburst activity.
However, alternative mechanisms, such as AGN or shock heating, could also contribute; current datasets render it extremely challenging to clearly quantify their contributions to the observed high EWs.

\item We found a unique overdensity of EELGs ($>1000$ \AA) at $z\sim7.3$.
The overdensity factor is more than 50 compared to the average field.
This region coincides with the spectroscopically confirmed overdensity at $z_{\mathrm{spec}}$ = 7.265 $\pm$ 0.007 with $\delta_{\mathrm{gal}}$ = 6.631 $\pm$ 1.322, and $\log_{10}(M_{\mathrm{Halo}}/\mathrm{M_{\odot}})$ = 11.683 $\pm$ 0.106 \citep{Helton2023}.
The presence of such a unique overdensity of EELGs is consistent with the idea that ionizing bubbles formed around high-density regions in the early Universe.

\end{enumerate}

\section*{Acknowledgement}
This work is based on observations made with the NASA/ESA/CSA James Webb Space Telescope. 
The data were obtained from the Mikulski Archive for Space Telescopes at the Space Telescope Science Institute, which is operated by the Association of Universities for Research in Astronomy, Inc., under NASA contract NAS 5-03127 for JWST. 
This work is supported by JST (the establishment of university fellowships towards the creation of science technology innovation) Grant Number JPMJFS2102, and Graduate Program on Physics for the Universe (GP-PU), Tohoku University.
TK acknowledges financial support from JSPS KAKENHI Grant Numbers 24H00002 (Specially Promoted Research by T. Kodama et al.) and 22K21349 (International Leading Research by S. Miyazaki et al.).
J.M.P.M. acknowledges funding from the European Union’s Horizon-Europe research and innovation programme under the Marie Skłodowska-Curie grant agreement No 101106626.

\bibliography{main.bib}

\begin{table}[h]
\begin{center}
\caption{Physical properties of \hb\ + \oiii\ emitters with spatially offsets.}
\begin{tabular}{ccccccccc}
\hline
ID & RA & DEC &Rest-frame EW$_{0}$ & $\log_{10}$($M_{\star}$) & $\Delta$MS& $d$(Line-UV) & $d$(Line-stellar)& $d$(UV-stellar)\\
& &&[\AA ] & [$\mathrm{M_{\odot}}$] & [dex] & [kpc] & [kpc] & [kpc]\\\hline\hline
1&34.32393 & $-$5.28589 & 3765 $\pm$ 270 & $7.42_{-0.26}^{+0.29}$ & $0.49_{-0.10}^{+0.14} $& 0.41 $\pm$ 0.04 & 0.43 $\pm$ 0.03 & 0.03 $\pm$ 0.03\\
2&34.28884 & $-$5.27660 & 1191 $\pm$ 62 & $8.57_{-0.09}^{+0.08}$ & $0.07_{-0.16}^{+0.15} $& 0.31 $\pm$ 0.03 & 0.43 $\pm$ 0.02 & 0.15 $\pm$ 0.03\\
3&34.48392 & $-$5.26479 & 2021 $\pm$ 127 & $8.19_{-0.23}^{+0.28}$ & $0.35_{-0.07}^{+0.19} $& 0.33 $\pm$ 0.02 & 0.30 $\pm$ 0.03 & 0.20 $\pm$ 0.02\\
4&34.42499 & $-$5.26518 &1804 $\pm$ 50 & $8.52_{-0.20}^{+0.15}$ & $0.15_{-0.14}^{+0.19} $& 0.38 $\pm$ 0.03 & 0.46 $\pm$ 0.03 & 0.08 $\pm$ 0.03\\
5&34.26580 & $-$5.22430 &741 $\pm$ 62 & $8.62_{-0.20}^{+0.19}$ & $0.36_{-0.18}^{+0.17} $& 0.38 $\pm$ 0.02 & 0.38 $\pm$ 0.02 & 0.12 $\pm$ 0.02\\
6&34.30996 & $-$5.20919 &1589 $\pm$ 120  & $8.43_{-0.36}^{+0.30}$ & $0.35_{-0.02}^{+0.02} $& 0.50 $\pm$ 0.03 & 0.49 $\pm$ 0.03 & 0.11 $\pm$ 0.03\\
7&34.31370 & $-$5.20309 &1279 $\pm$ 53 & $8.81_{-0.09}^{+0.16}$ & $0.66_{-0.17}^{+0.15} $& 0.51 $\pm$ 0.03 & 0.53 $\pm$ 0.03 & 0.23 $\pm$ 0.03\\
8&53.19312 & $-$27.77004 &844 $\pm$ 20  & $7.83_{-0.11}^{+0.09}$ & $-0.09_{-0.06}^{+0.10}$ & 0.36 $\pm$ 0.03 & 0.35 $\pm$ 0.03 & 0.08 $\pm$ 0.03\\
9&53.14942 & $-$27.76099 &1983 $\pm$ 380 & $7.07_{-0.10}^{+0.35}$ & $0.46_{-0.03}^{+0.03}$& 0.45 $\pm$ 0.03 & 0.40 $\pm$ 0.03 & 0.06 $\pm$ 0.02\\
10 &  53.18245 & $-$27.74478 & 1274 $\pm$ 82 & $7.25_{-0.14}^{+0.19}$ & $0.28_{-0.04}^{+0.03}$& 0.34 $\pm$ 0.02 & 0.44 $\pm$ 0.02 & 0.12 $\pm$ 0.02\\
11& 150.10645 & 2.27803 & 571 $\pm$ 109 & $9.29_{-0.13}^{+0.12}$ & $-0.29_{-0.15}^{+0.18}$ & 0.66 $\pm$ 0.03 & 0.66 $\pm$ 0.03 & 0.16 $\pm$ 0.03\\
12& 150.18289 & 2.45999& 2418 $\pm$ 287 & $8.26_{-0.11}^{+0.14}$ & $0.65_{-0.14}^{+0.13} $& 0.73 $\pm$ 0.03 & 0.92 $\pm$ 0.03 & 0.22 $\pm$ 0.03\\
13& 260.75092 & 65.82269 & 791 $\pm$ 61 & $8.26_{-0.11}^{+0.16}$ & $-0.07_{-0.14}^{+0.16}$ & 0.39 $\pm$ 0.03 & 0.34 $\pm$ 0.02 & 0.08 $\pm$ 0.03\\\hline
\end{tabular}
\label{tab:offset}
\end{center}
\end{table}

\begin{table}
\begin{center}
\caption{Catalog of \hb\ + \oiii\ emitters with rest-frame \hb\ + \oiii\ EW $>$ 3000 \AA.}
\label{tab:HEW}
\begin{tabular}{ccccccc}\hline
\quad ID \quad & \quad  RA \quad & \quad  DEC \quad  &\quad  Rest-frame EW$_{0}$ \quad & \quad  $\log_{10}$($M_{\star}$)\quad  & \quad $\Delta$MS\quad &\quad $\beta$ \quad \\
&&& \quad [\AA ]\quad  & \quad [$\mathrm{M_{\odot}}$] \quad & \quad [dex]\quad  \\\hline\hline
\quad HEW1 \quad & \quad  177.45447 \quad & \quad 22.47224 &\quad  5251 $\pm$ 532 \quad & \quad $7.24_{-0.14}^{+0.15}$\quad  & \quad $0.58_{-0.03}^{+0.04}$\quad  &\quad $-2.45_{-0.09}^{+0.04}$ \quad\\
\quad HEW2 \quad & \quad 260.70346 \quad & \quad 65.77679 \quad &\quad  3236 $\pm$ 501 \quad & \quad $7.63_{-0.13}^{+0.39}$ \quad & \quad $0.32_{-0.09}^{+0.22}$\quad &\quad $-2.20_{-0.18}^{+0.22}$  \quad\\
\quad HEW3 \quad & \quad 260.78864 \quad & \quad 65.80657 \quad &\quad  3695 $\pm$ 297 \quad & \quad $7.34_{-0.08}^{+0.09}$ \quad & \quad $0.70_{-0.03}^{+0.03}$\quad  &\quad $-2.68_{-0.03}^{+0.03}$\quad\\
\quad HEW4 \quad & \quad 150.17801 \quad & \quad 2.23936 \quad &\quad  3195 $\pm$ 439 \quad & \quad $7.49_{-0.09}^{+0.22}$ \quad & \quad $0.70_{-0.03}^{+0.03}$\quad  &\quad $-2.70_{-0.04}^{+0.03}$\quad\\
\quad HEW5 \quad & \quad 150.11996 \quad & \quad 2.28068 \quad &\quad  4811 $\pm$ 738 \quad & \quad $7.50_{-0.38}^{+0.31}$ \quad & \quad $0.54_{-0.16}^{+0.28}$\quad  &\quad $-2.17_{-0.20}^{+0.28}$\quad\\
\quad HEW6 \quad & \quad 150.18399 \quad & \quad 2.28490 \quad &\quad  3426 $\pm$ 154 \quad & \quad $7.96_{-0.58}^{+0.62}$ \quad & \quad $0.48_{-0.26}^{+0.31}$\quad  &\quad $-1.60_{-0.20}^{+0.31}$\quad\\
\quad HEW7 \quad & \quad 150.11908 \quad & \quad 2.41567 \quad &\quad  3254 $\pm$ 481 \quad & \quad $8.22_{-0.30}^{+0.22}$ \quad & \quad $0.40_{-0.28}^{+0.37}$\quad  &\quad $-1.89_{-0.25}^{+0.37}$\quad\\
\quad HEW8 \quad & \quad 34.34293 \quad & \quad $-$5.30944 \quad &\quad  3531 $\pm$ 441 \quad & \quad $7.33_{-0.41}^{+0.46}$ \quad & \quad $0.43_{-0.18}^{+0.33}$\quad  &\quad $-2.15_{-0.22}^{+0.33}$\quad\\
\quad HEW9 \quad & \quad 34.52908 \quad & \quad $-$5.29681 \quad &\quad  3248 $\pm$ 380 \quad & \quad $7.86_{-0.17}^{+0.21}$ \quad & \quad $0.48_{-0.05}^{+0.09}$\quad  &\quad $-2.29_{-0.14}^{+0.09}$\quad \\
\quad HEW10 \quad & \quad 34.32393 \quad  & \quad $-$5.28589 \quad &\quad  3765 $\pm$ 682 \quad & \quad $7.42_{-0.26}^{+0.29}$ \quad & \quad $0.49_{-0.10}^{+0.14}$\quad  &\quad $-2.39_{-0.18}^{+0.14}$\quad  \\
\quad HEW11 \quad & \quad 34.50491 \quad & \quad $-$5.25796 \quad &\quad  3605 $\pm$ 445 \quad & \quad $8.12_{-0.21}^{+0.22}$ \quad & \quad $0.56_{-0.26}^{+0.34}$ \quad &\quad $-1.72_{-0.24}^{+0.34}$\quad\\
\quad HEW12 \quad & \quad 34.28159 \quad & \quad $-$5.23756 \quad &\quad  4152 $\pm$ 253 \quad & \quad $8.26_{-0.43}^{+0.26}$ \quad & \quad $0.64_{-0.20}^{+0.23}$\quad  &\quad $-1.78_{-0.17}^{+0.23}$\quad\\
\quad HEW13 \quad & \quad 34.39332 \quad & \quad $-$5.23562 \quad &\quad  5571 $\pm$ 570 \quad & \quad $7.53_{-0.51}^{+0.31}$ \quad & \quad $0.69_{-0.25}^{+0.27}$\quad  &\quad $-1.96_{-0.21}^{+0.27}$\quad\\
\quad HEW14 \quad & \quad 34.45022 \quad & \quad $-$5.21322 \quad &\quad  6848 $\pm$ 840 \quad & \quad $7.05_{-0.53}^{+0.31}$ \quad & \quad $0.79_{-0.10}^{+0.18}$\quad  &\quad $-2.34_{-0.11}^{+0.18}$\quad\\
\quad HEW15 \quad & \quad 34.49124 \quad & \quad $-$5.21223 \quad &\quad  4274 $\pm$ 504 \quad & \quad $7.66_{-0.13}^{+0.24}$ \quad & \quad $0.63_{-0.06}^{+0.06}$\quad  &\quad $-2.45_{-0.15}^{+0.06}$\quad\\
\quad HEW16 \quad & \quad 53.16639 \quad & \quad $-$27.81237 \quad &\quad  3255 $\pm$ 319 \quad & \quad $6.95_{-0.22}^{+0.31}$ \quad & \quad $0.37_{-0.05}^{+0.05}$\quad  & \quad$-2.41_{-0.16}^{+0.05}$\quad\\
\quad HEW17 \quad & \quad 53.17685 \quad & \quad $-$27.79679 \quad &\quad  3238 $\pm$ 601 \quad & \quad $7.25_{-0.17}^{+0.25}$ \quad & \quad $0.25_{-0.05}^{+0.19}$\quad  & \quad$-2.22_{-0.22}^{+0.19}$\quad\\
\quad HEW18 \quad & \quad 53.18224 \quad & \quad $-$27.77183 \quad &\quad  3104 $\pm$ 135 \quad & \quad $7.08_{-0.12}^{+0.26}$ \quad & \quad $0.31_{-0.03}^{+0.09}$\quad   &\quad $-2.27_{-0.11}^{+0.09}$\quad\\
\quad HEW19 \quad & \quad 53.16869 \quad & \quad $-$27.73720 \quad &\quad  3769 $\pm$ 115 \quad & \quad $7.35_{-0.12}^{+0.20}$ \quad & \quad $0.55_{-0.04}^{+0.02}$\quad  & \quad$-2.47_{-0.06}^{+0.02}$\quad\\\hline
\end{tabular}
\end{center}
\end{table}

\appendix
\renewcommand{\thefigure}{A\arabic{figure}}
\setcounter{figure}{0}
\section{Comparison of Photometric and Spectroscopic Redshifts}

In this appendix, we provide a comprehensive analysis comparing the photometric redshifts (photo-$z$) used in our study with available spectroscopic redshifts (spec-$z$). 
The photo-$z$ estimates are critical in two parts of our analysis: first, in the selection of emission-line galaxies via their probability distribution functions, and second, in the SED fitting process.

\begin{figure*}
\begin{center} 
\includegraphics[width=82.5mm]{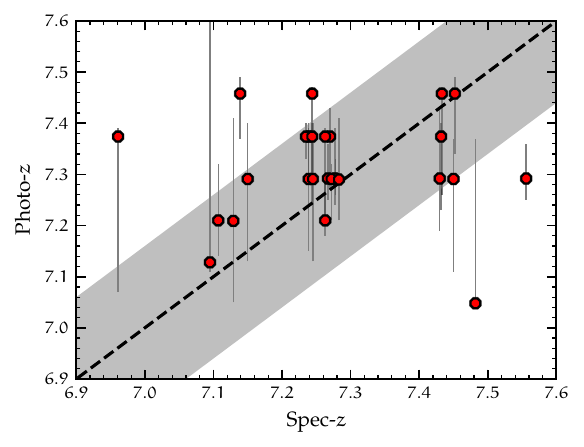}
\caption{Comparison between photometric redshifts (photo-$z$) and spectroscopic redshifts (spec-$z$) for 27 objects in the UDS and GOODS-S fields. All objects fall within the expected redshift range determined by the filter response functions ($6.6<z<7.5$). The variability in each dataset is represented by a standard deviation of 0.16, indicated by the shaded region.}
\label{fig:photoz_specz}
\end{center} 
\end{figure*}

Data from the JADES and Ruby \citep[JWST-GO-4233; PI de Graaff \& Brammer: e.g.,][]{deGraaff2024} surveys were used to assess our selection criteria in three well-studied fields: COSMOS, UDS, and GOODS-S. 
We identify a total of 28 objects with confirmed spec-$z$ measurements. 
Figure~\ref{fig:photoz_specz} shows the comparison between photo-$z$ and spec-$z$ for these objects, with the dashed line indicating the one-to-one relation.

Our analysis indicates that all objects lie within the redshift range predicted by the filter response functions. Despite the intrinsic selection biases of each survey, the agreement between photo-$z$ and spec-$z$ confirms that our methods reliably select high-$z$ galaxies up to $z\sim7$. In particular, approximately 84\% of the objects show a photo-$z$ accuracy within $\pm0.3$, underscoring the robustness of the photo-$z$ estimates employed for both the initial selection of emission-line galaxies and the subsequent SED fitting.

To mitigate contamination, our methodology combines the use of color-color diagrams, rigorous color selection criteria, and the full probability distributions from the photo-$z$ analysis. Although we cannot directly quantify the overall contamination rate, the strong correlation between photo-$z$ and spec-$z$ values suggests that our sample is largely uncontaminated and suitable for studies of high-redshift galaxy populations.

In summary, the detailed comparison presented here validates the photo-$z$ measurements used in our work and supports the reliability of our emission-line galaxy selection strategy. 
This MB approach proves to be effective in advancing our understanding of galaxy evolution in the early Universe and will serve as a valuable guideline for future deep surveys.

\renewcommand{\thefigure}{B\arabic{figure}}
\setcounter{figure}{0}

\begin{figure*}
    \begin{center}
    \includegraphics[width=164mm]{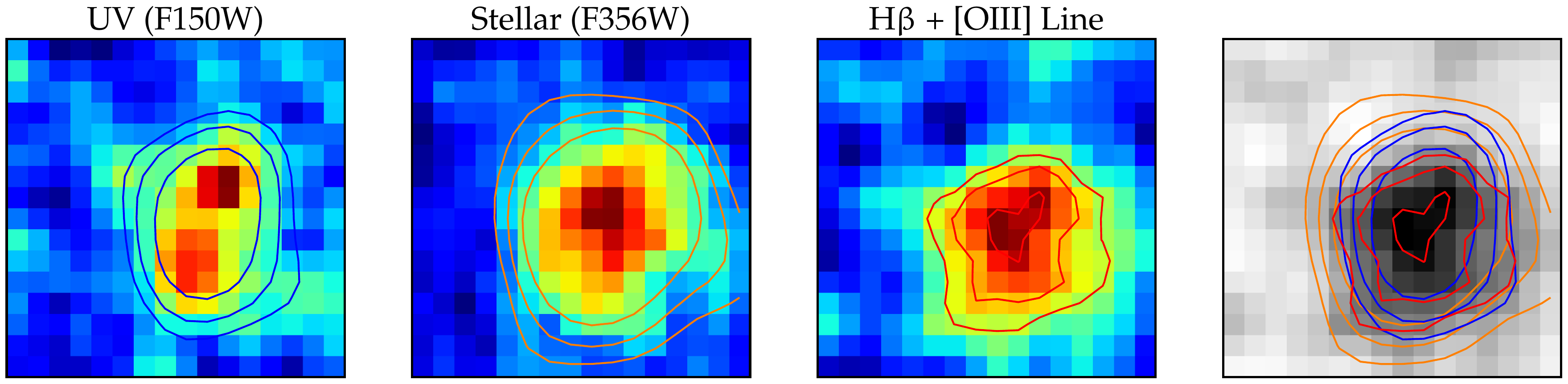}
    \includegraphics[width=164mm]{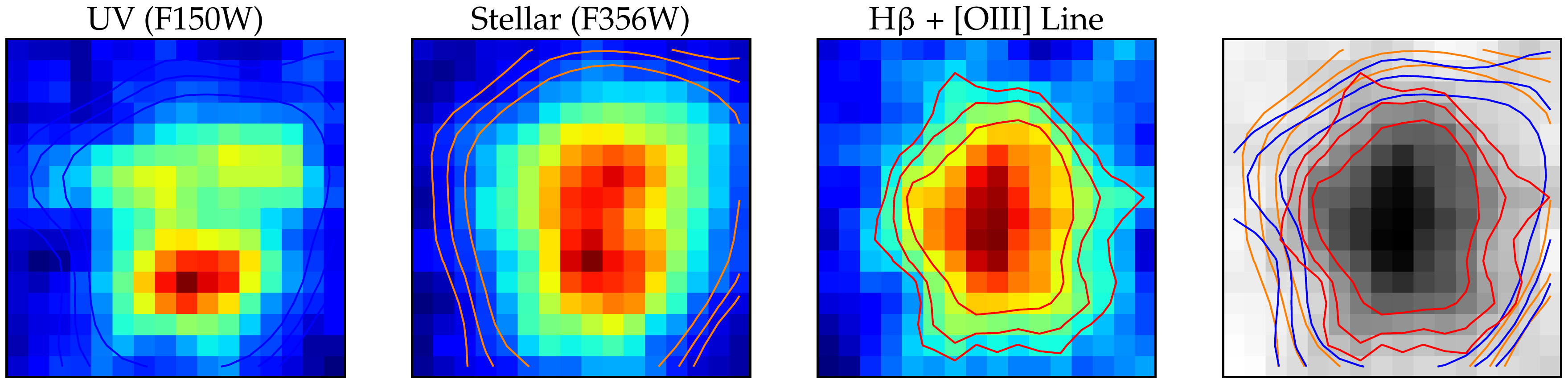}
    \includegraphics[width=164mm]{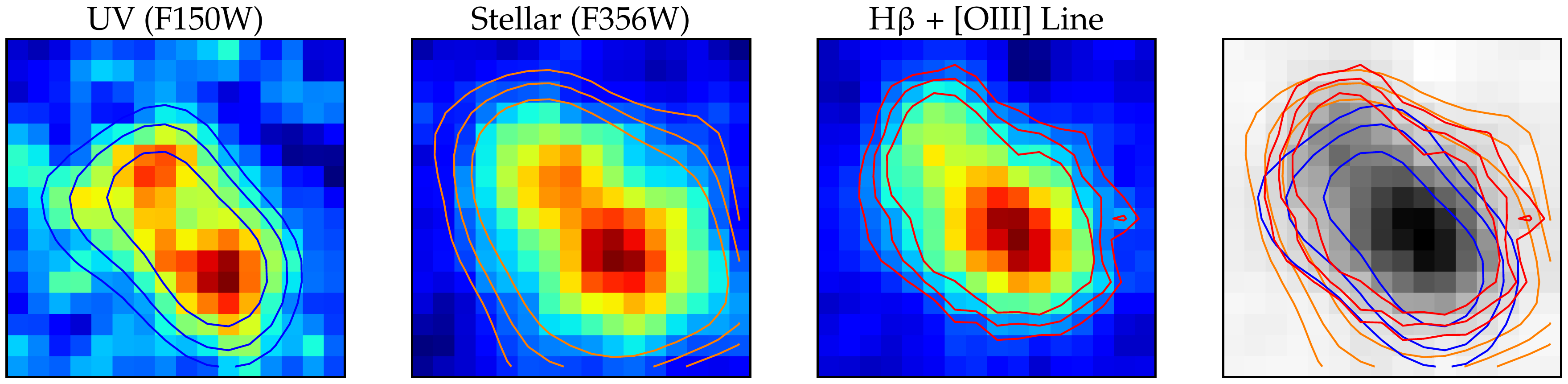}
    \includegraphics[width=164mm]{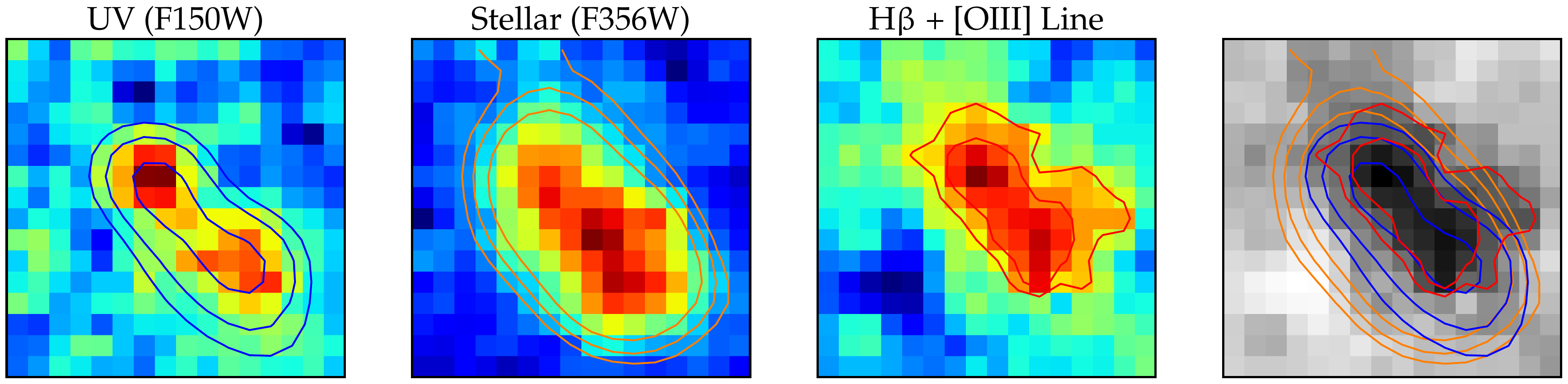}
    \includegraphics[width=164mm]{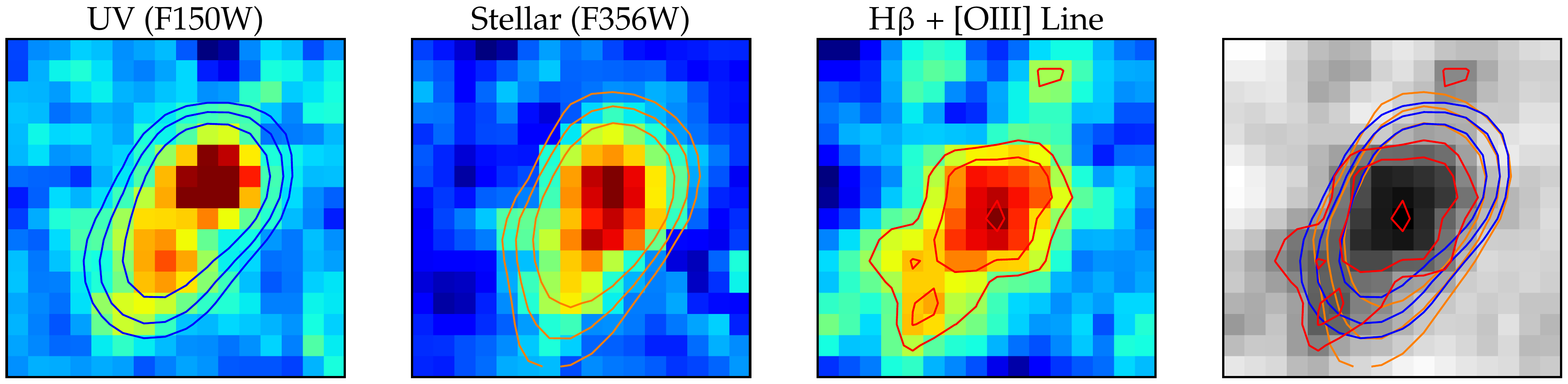}
    \caption{Examples of clumpy emitters identified rest-frame UV band (F150W). }
    \label{fig:clumpy}
    \end{center}
\end{figure*}

\begin{figure*}[h]
\begin{minipage}{0.5\hsize} 
\begin{center} 
\includegraphics[width=82.5mm]{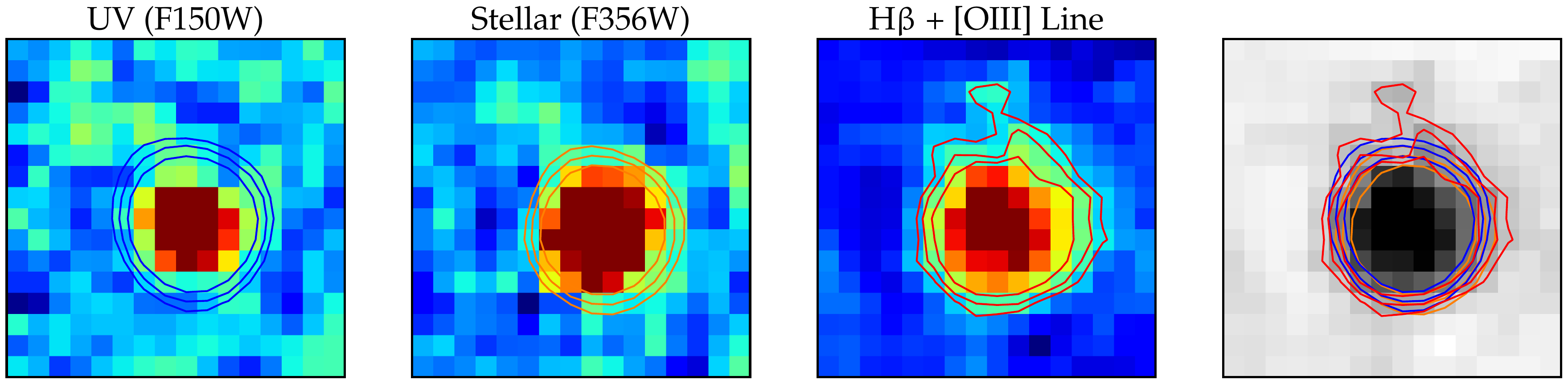}
\includegraphics[width=82.5mm]{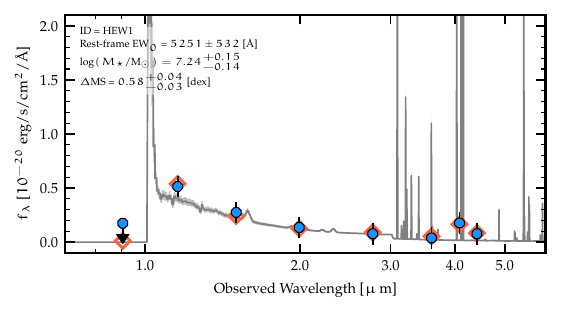}
\end{center} 
\end{minipage} 
\begin{minipage}{0.5\hsize} 
\begin{center} 
\includegraphics[width=82.5mm]{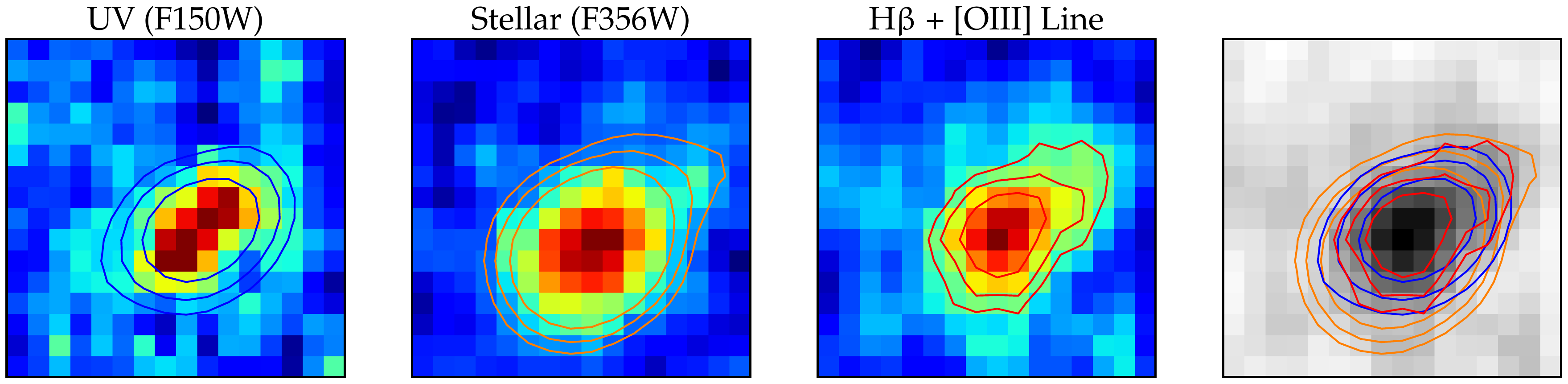}
\includegraphics[width=82.5mm]{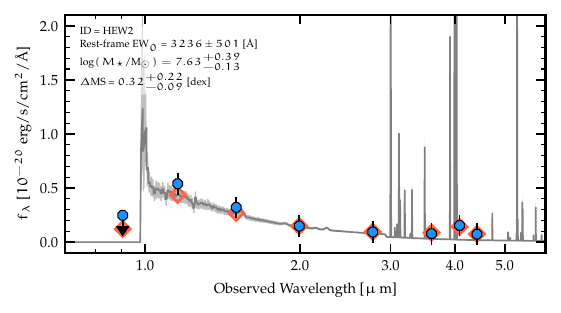}
\end{center} 
\end{minipage}
\caption{(Top) Image cutouts of HEW1 and HWE2. The images show the UV component (F150W), stellar component (F356W; rest-frame optical), and \hb+\oiii\ emission line distributions from the left to the right panels. The contours show 5, 7, and 10 $\sigma$. In the right most panel, the emission line distribution (red contours and the underlying image) is clearly offset from the UV (blue contours) and stellar mass component (yellow contours). Box size is $2^{\prime\prime}.0$, corresponding to around 10 kpc. (Bottom) An illustrative SED of individual HEWs. Blue-filled circles show the observed photometric fluxes (F090W, F150W, F115W, F200W, F277W, F356W, F410M, and F444W). The open diamonds show the model fluxes for the corresponding input filters. The grey line and the grey zone around it show the best-fit SED and the $1\,\sigma$ uncertainty range, respectively.}
\label{fig:ImageHEWsSEDs_s}
\end{figure*}

\begin{figure*}
\begin{minipage}{0.5\hsize} 
\begin{center} 
\includegraphics[width=82.5mm]{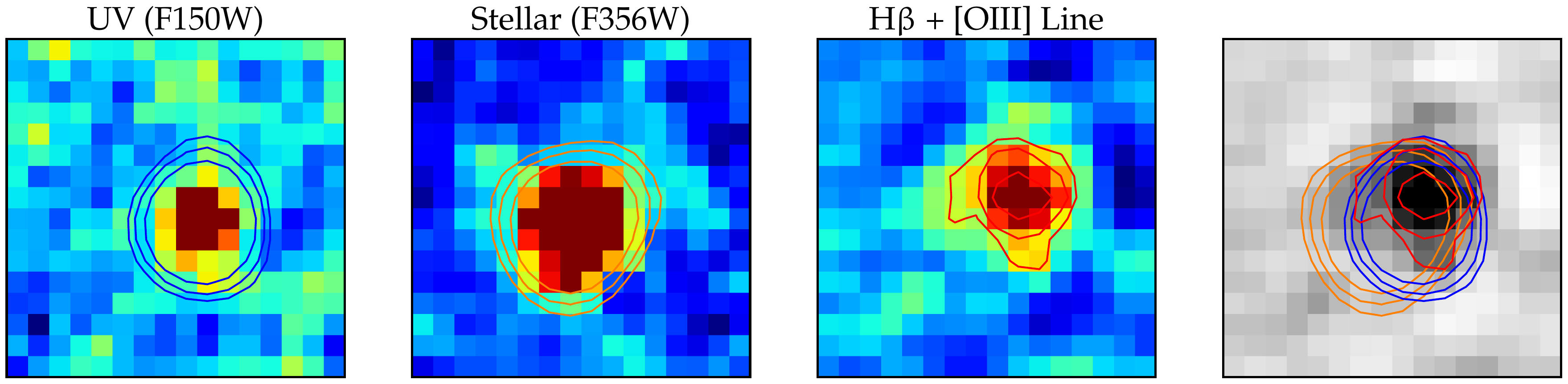}
\includegraphics[width=82.5mm]{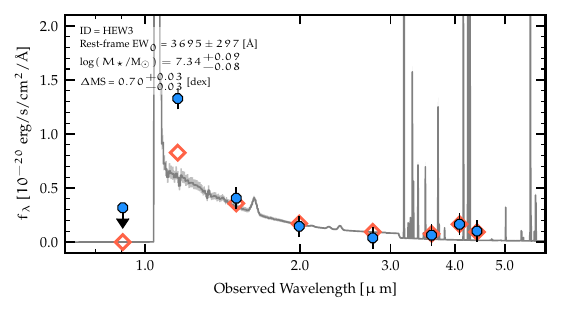}
\end{center} 
\end{minipage} 
\begin{minipage}{0.5\hsize} 
\begin{center} 
\includegraphics[width=82.5mm]{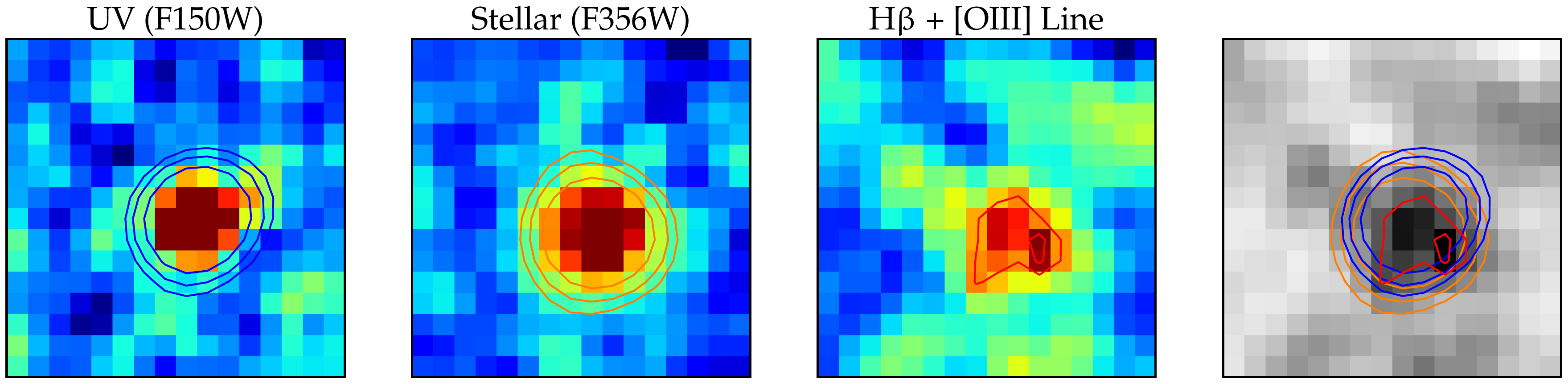}
\includegraphics[width=82.5mm]{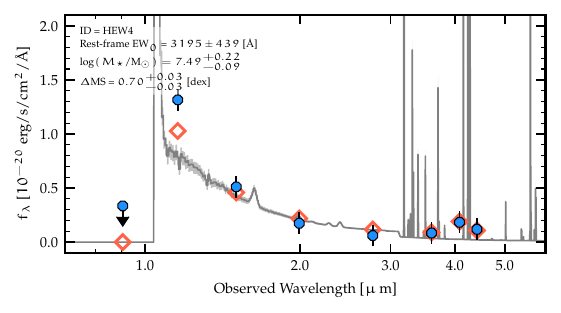}
\end{center} 
\end{minipage}
\caption{(Continued) Image cutouts and SEDs of HEW3 and HWE4.}
\vspace{+0.5cm}
\begin{minipage}{0.5\hsize} 
\begin{center} 
\includegraphics[width=82.5mm]{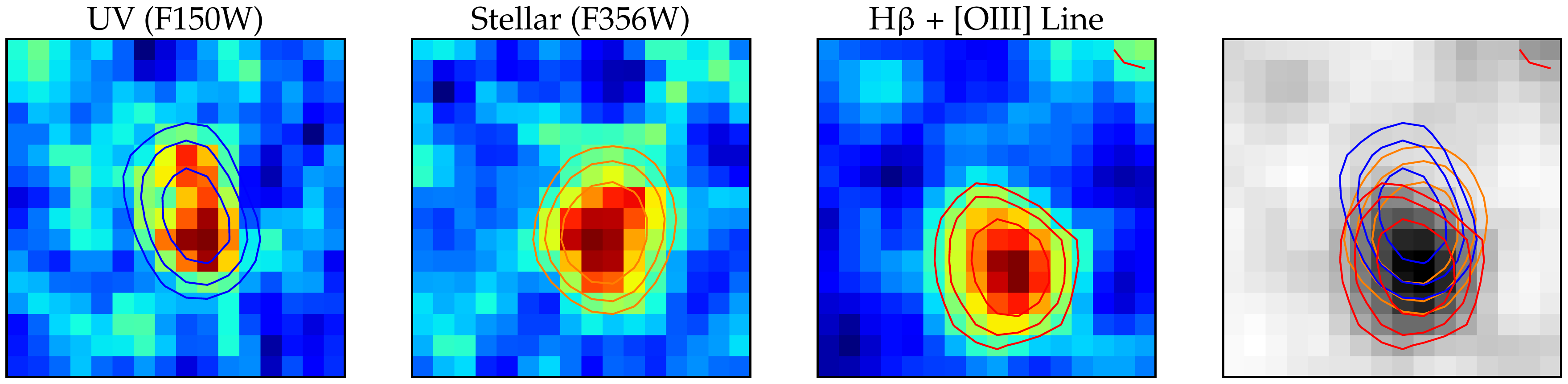}
\includegraphics[width=82.5mm]{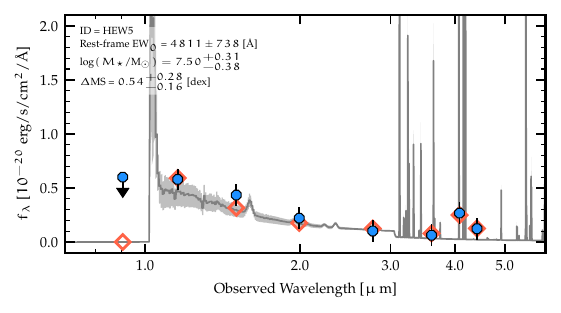}
\end{center} 
\end{minipage} 
\begin{minipage}{0.5\hsize} 
\begin{center} 
\includegraphics[width=82.5mm]{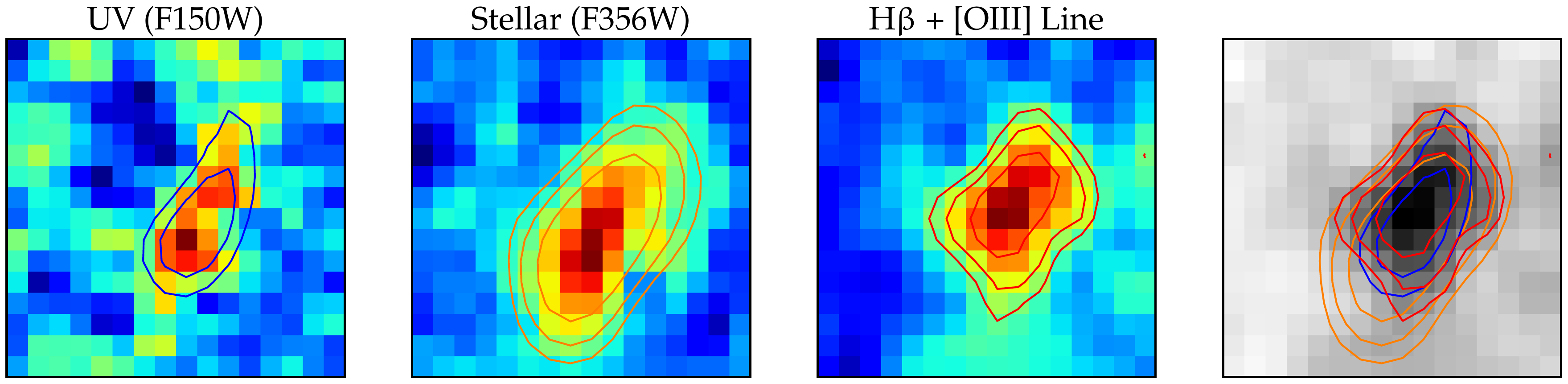}
\includegraphics[width=82.5mm]{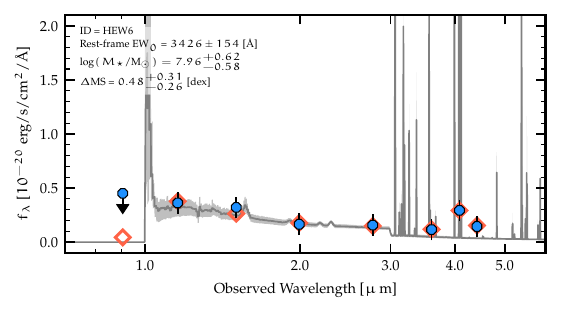}
\end{center} 
\end{minipage}
\caption{(Continued) Image cutouts and SEDs of HEW5 and HWE6.}
\vspace{+0.5cm}
\begin{minipage}{0.5\hsize} 
\begin{center} 
\includegraphics[width=82.5mm]{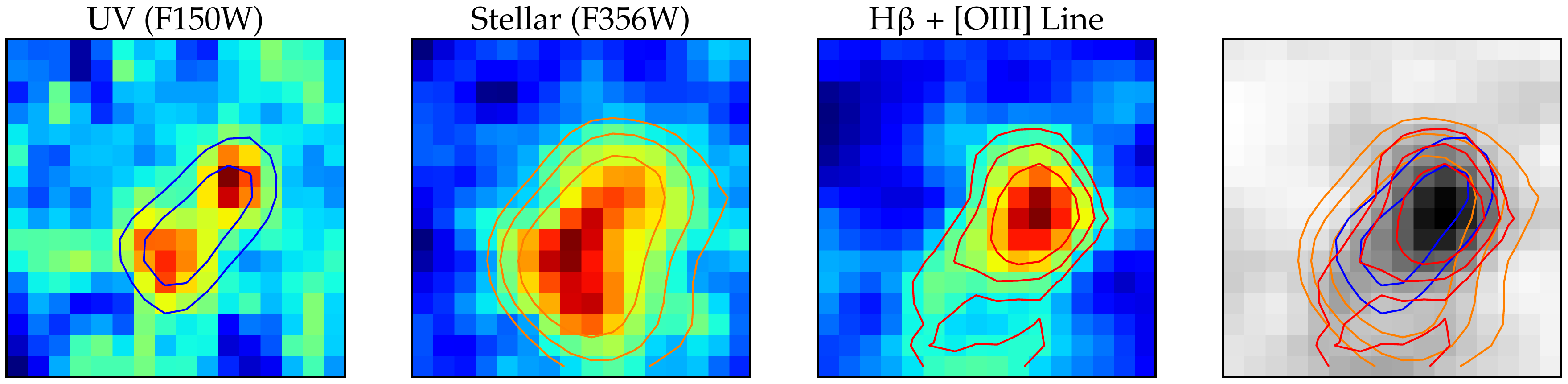}
\includegraphics[width=82.5mm]{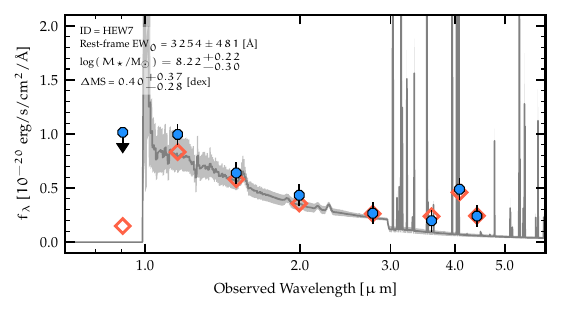}
\end{center} 
\end{minipage} 
\begin{minipage}{0.5\hsize} 
\begin{center} 
\includegraphics[width=82.5mm]{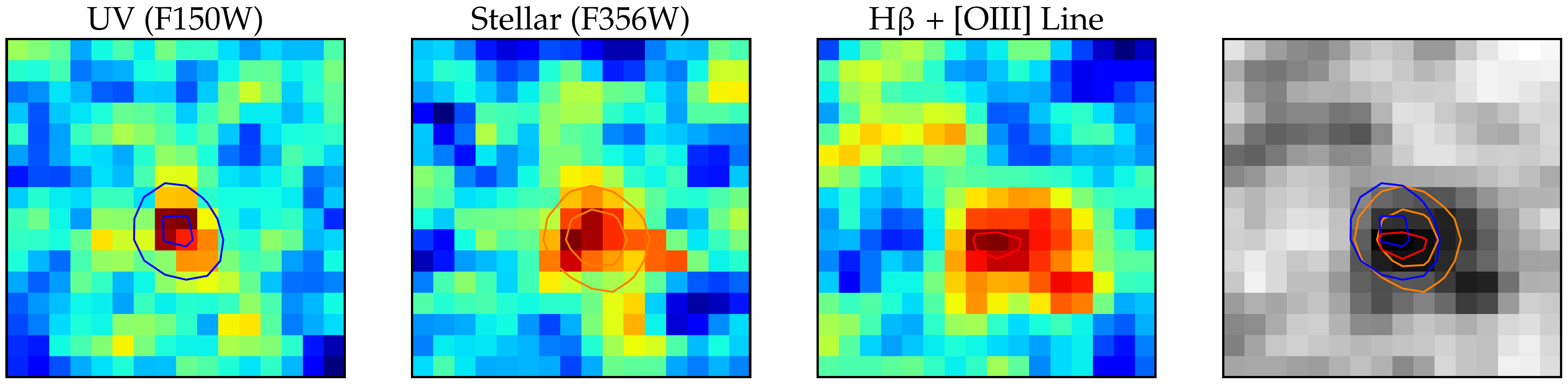}
\includegraphics[width=82.5mm]{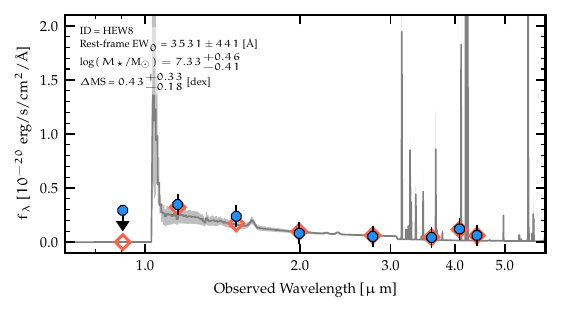}
\end{center} 
\end{minipage}
\caption{(Continued) Image cutouts and SEDs of HEW7 and HWE8.}
\end{figure*}

\begin{figure*}
\begin{minipage}{0.5\hsize} 
\begin{center} 
\includegraphics[width=82.5mm]{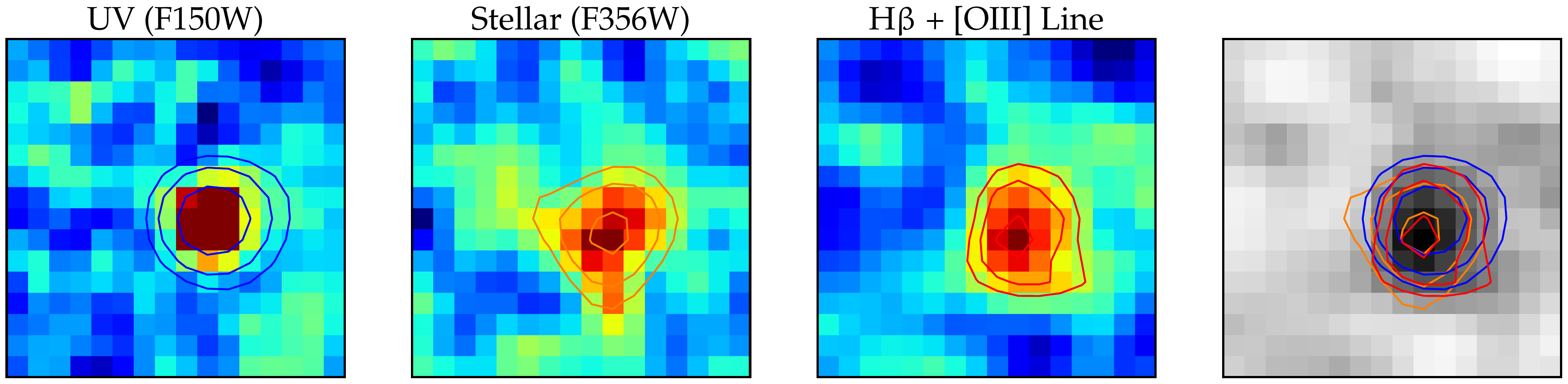}
\includegraphics[width=82.5mm]{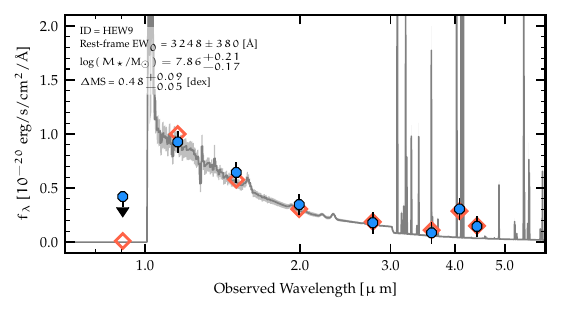}
\end{center} 
\end{minipage} 
\begin{minipage}{0.5\hsize} 
\begin{center} 
\includegraphics[width=82.5mm]{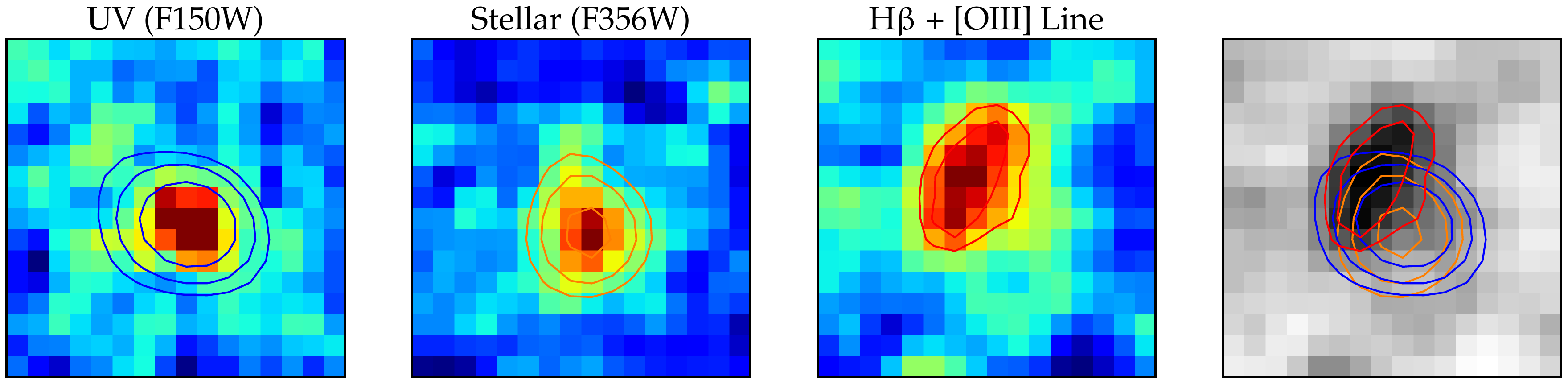}
\includegraphics[width=82.5mm]{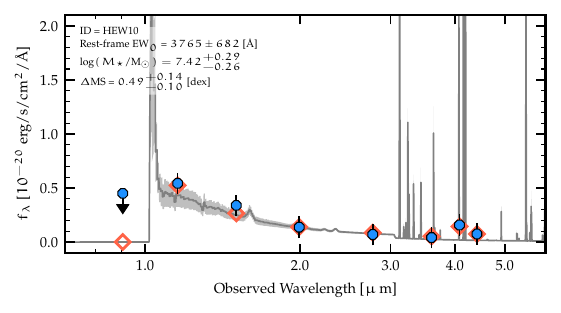}
\end{center} 
\end{minipage}
\caption{(Continued) Image cutouts and SEDs of HEW9 and HWE10.}
\vspace{+0.5cm}
\begin{minipage}{0.5\hsize} 
\begin{center} 
\includegraphics[width=82.5mm]{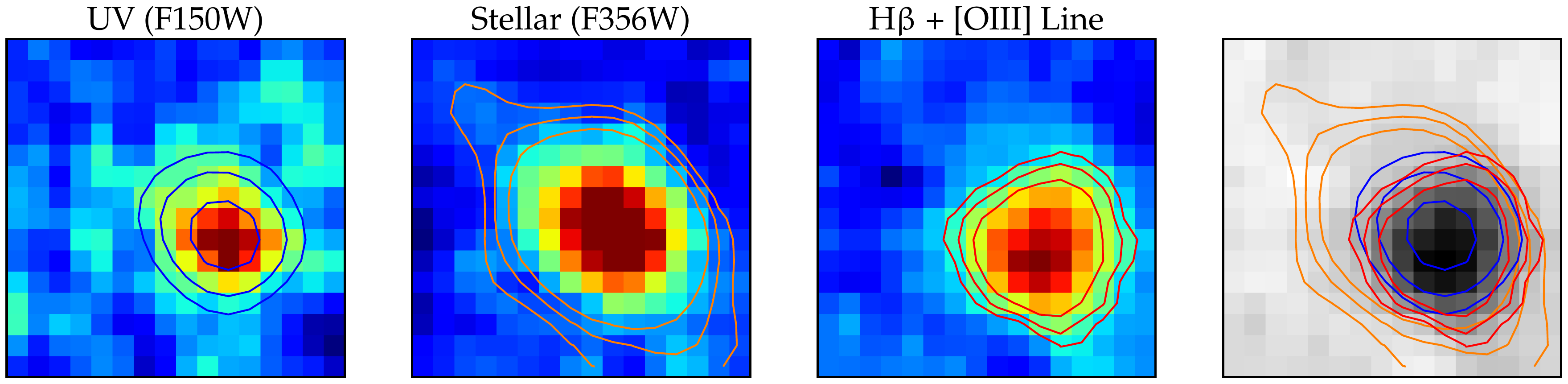}
\includegraphics[width=82.5mm]{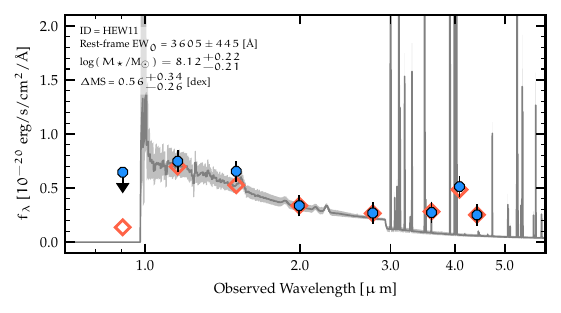}
\end{center} 
\end{minipage} 
\begin{minipage}{0.5\hsize} 
\begin{center} 
\includegraphics[width=82.5mm]{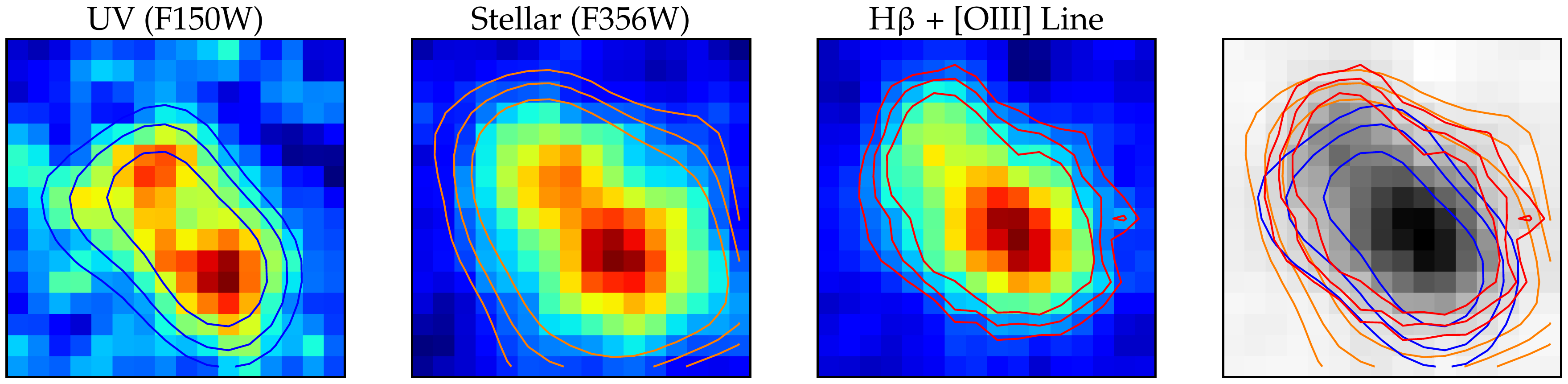}
\includegraphics[width=82.5mm]{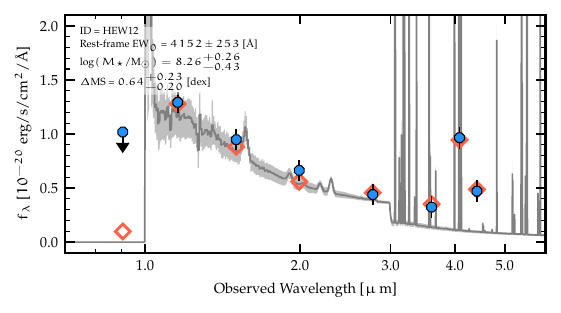}
\end{center} 
\end{minipage}
\caption{(Continued) Image cutouts and SEDs of HEW11 and HWE12.}
\vspace{+0.5cm}
\begin{minipage}{0.5\hsize} 
\begin{center} 
\includegraphics[width=82.5mm]{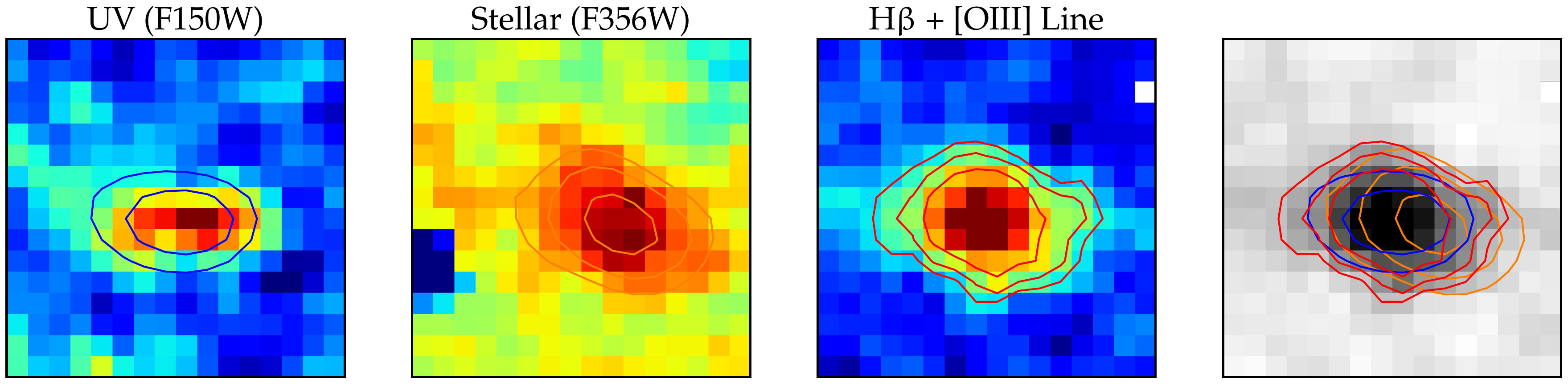}
\includegraphics[width=82.5mm]{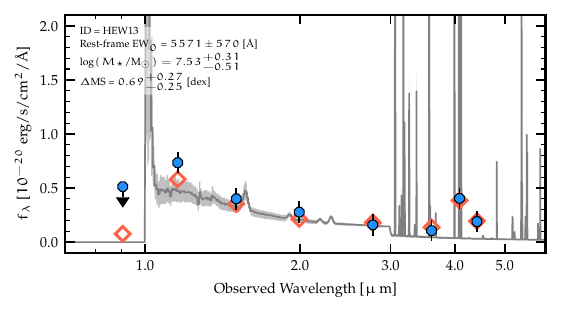}
\end{center} 
\end{minipage} 
\begin{minipage}{0.5\hsize} 
\begin{center} 
\includegraphics[width=82.5mm]{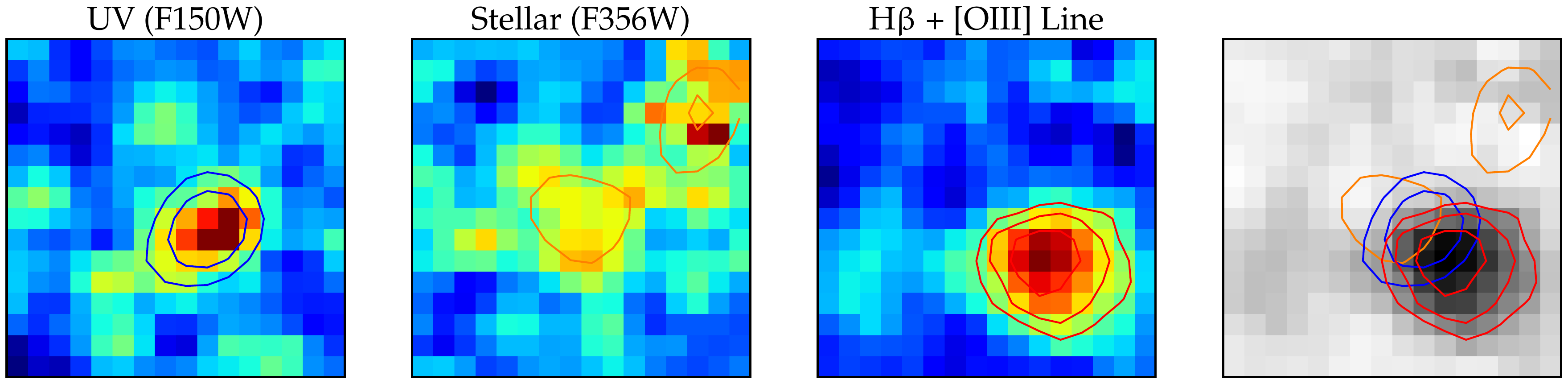}
\includegraphics[width=82.5mm]{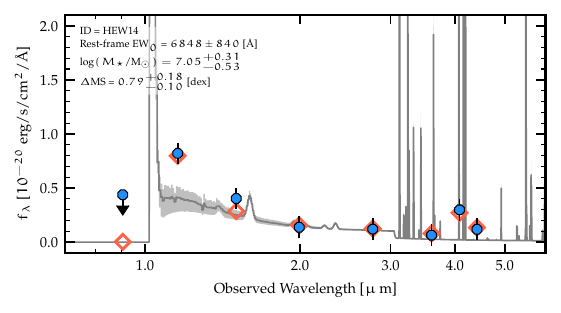}
\end{center} 
\end{minipage}
\caption{(Continued) Image cutouts and SEDs of HEW13 and HWE14.}
\end{figure*}

\begin{figure*}
\begin{minipage}{0.5\hsize} 
\begin{center} 
\includegraphics[width=82.5mm]{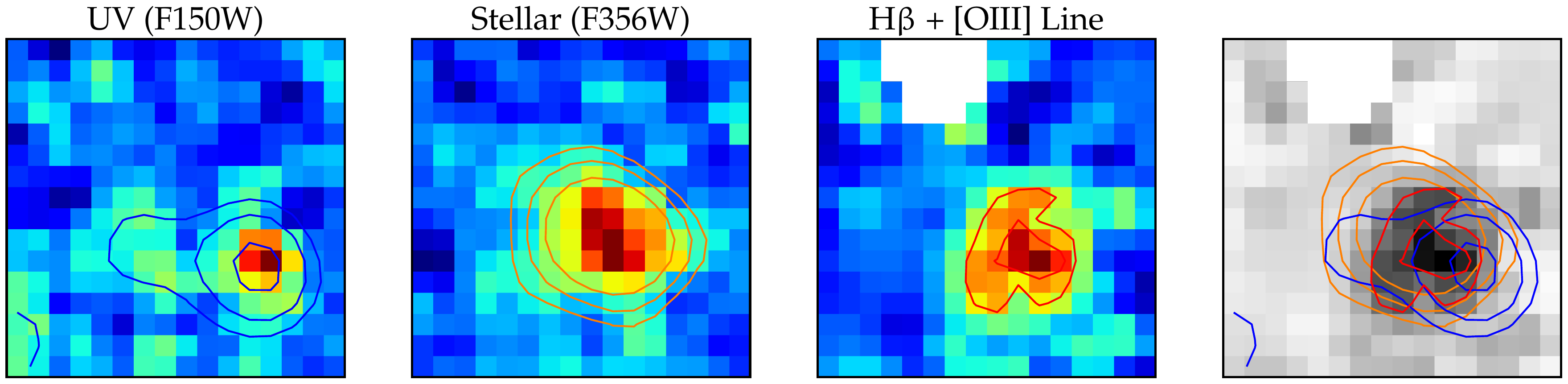}
\includegraphics[width=82.5mm]{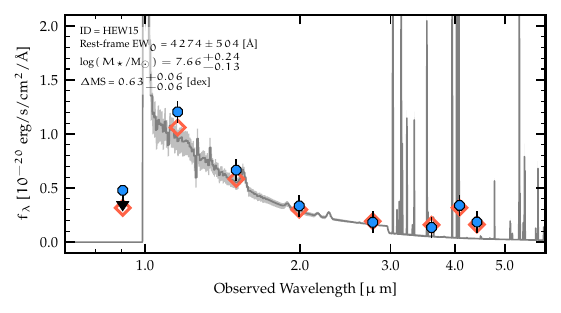}
\end{center} 
\end{minipage} 
\begin{minipage}{0.5\hsize} 
\begin{center} 
\includegraphics[width=82.5mm]{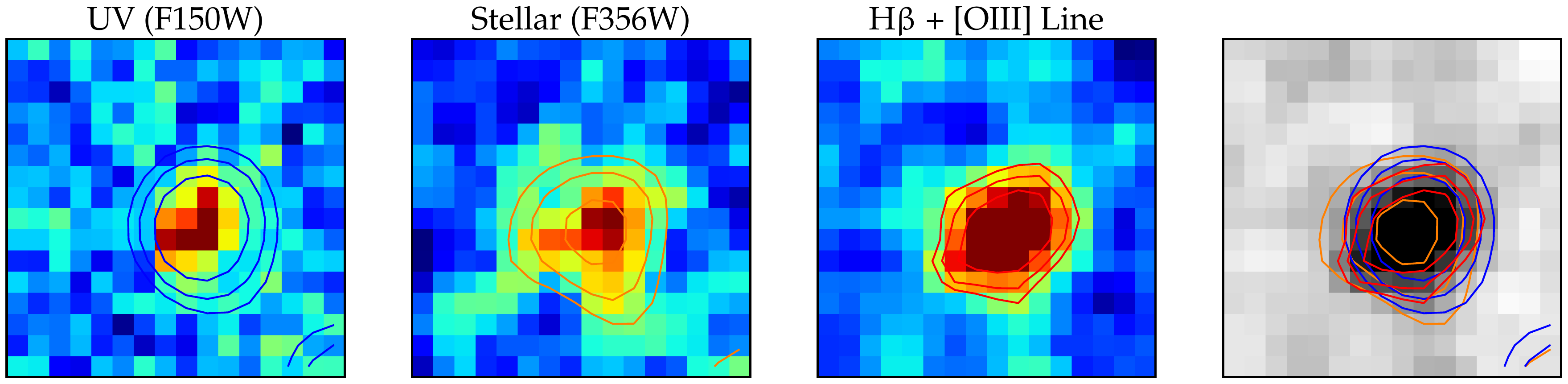}
\includegraphics[width=82.5mm]{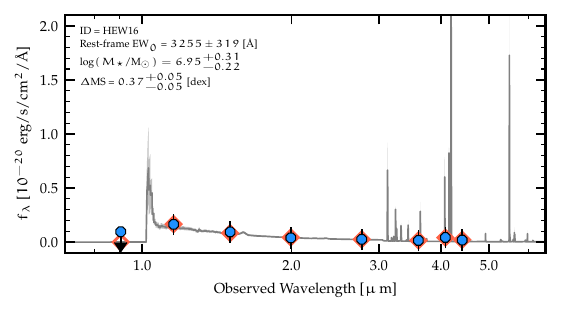}
\end{center} 
\end{minipage}
\caption{(Continued) Image cutouts and SEDs of HEW15 and HWE16.}
\vspace{+0.5cm}
\begin{minipage}{0.5\hsize} 
\begin{center} 
\includegraphics[width=82.5mm]{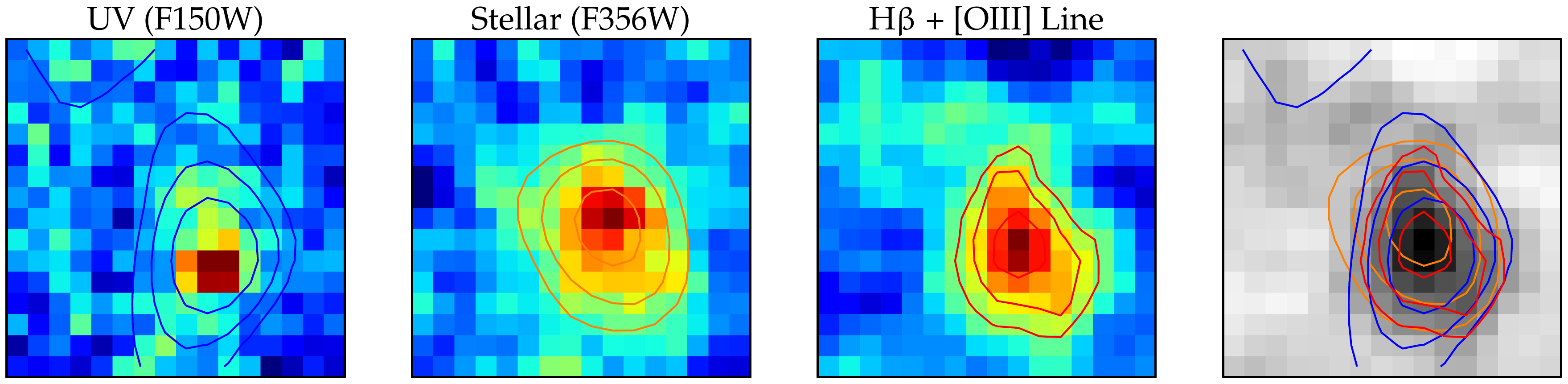}
\includegraphics[width=82.5mm]{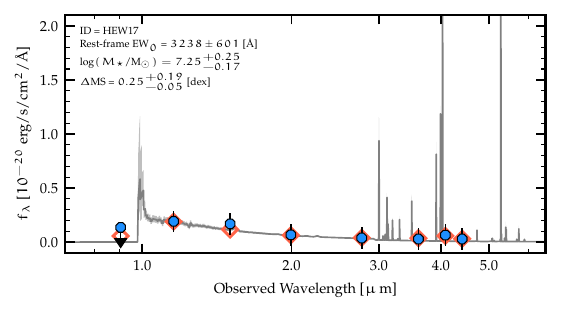}
\end{center} 
\end{minipage} 
\begin{minipage}{0.5\hsize} 
\begin{center} 
\includegraphics[width=82.5mm]{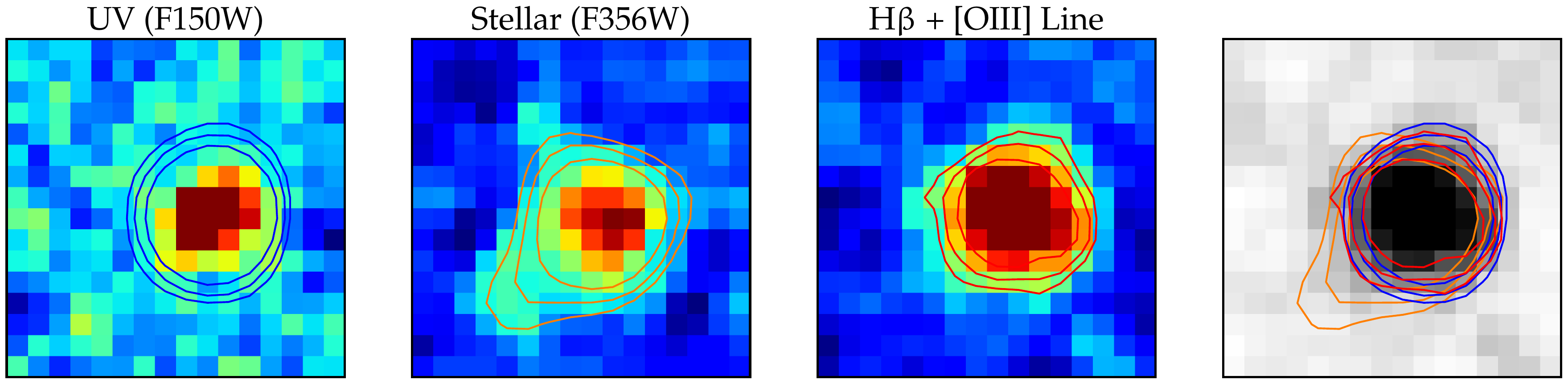}
\includegraphics[width=82.5mm]{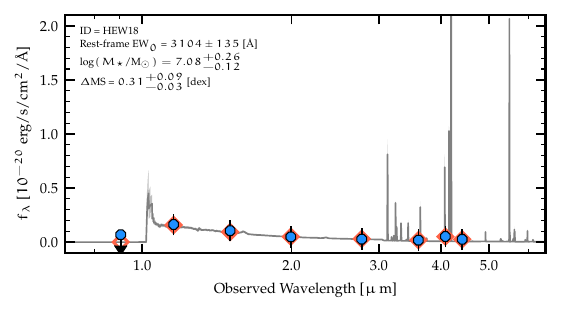}
\end{center} 
\end{minipage}
\caption{(Continued) Image cutouts and SEDs of HEW17 and HWE18.}
\vspace{+0.5cm}
\begin{minipage}{0.5\hsize} 
\begin{center} 
\includegraphics[width=82.5mm]{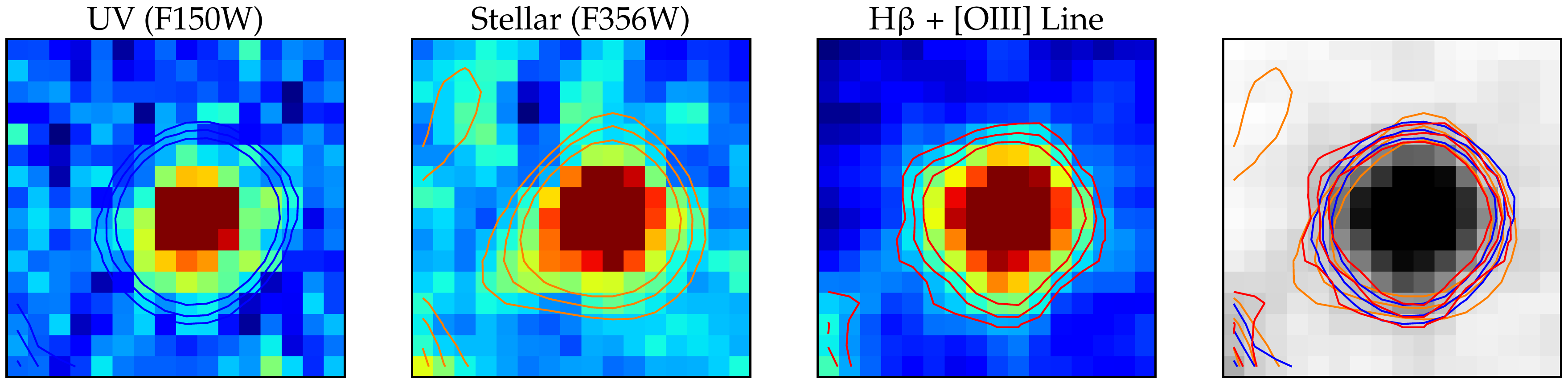}
\includegraphics[width=82.5mm]{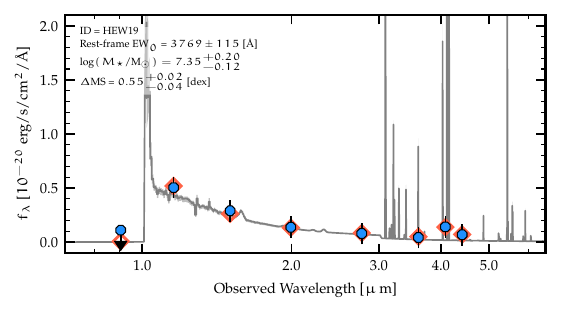}
\caption{(Continued) Image cutouts and SED of HEW19.}
\label{fig:ImageHEWsSEDs_e}
\end{center} 
\end{minipage}

\end{figure*}

\begin{figure*}
    \centering
    \includegraphics[width=168mm]{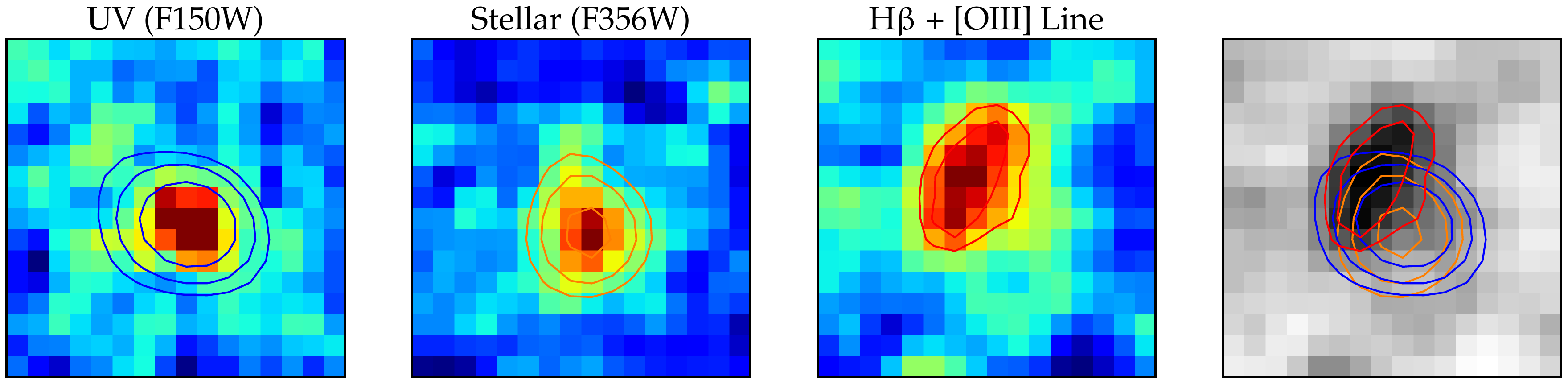}
    \includegraphics[width=168mm]{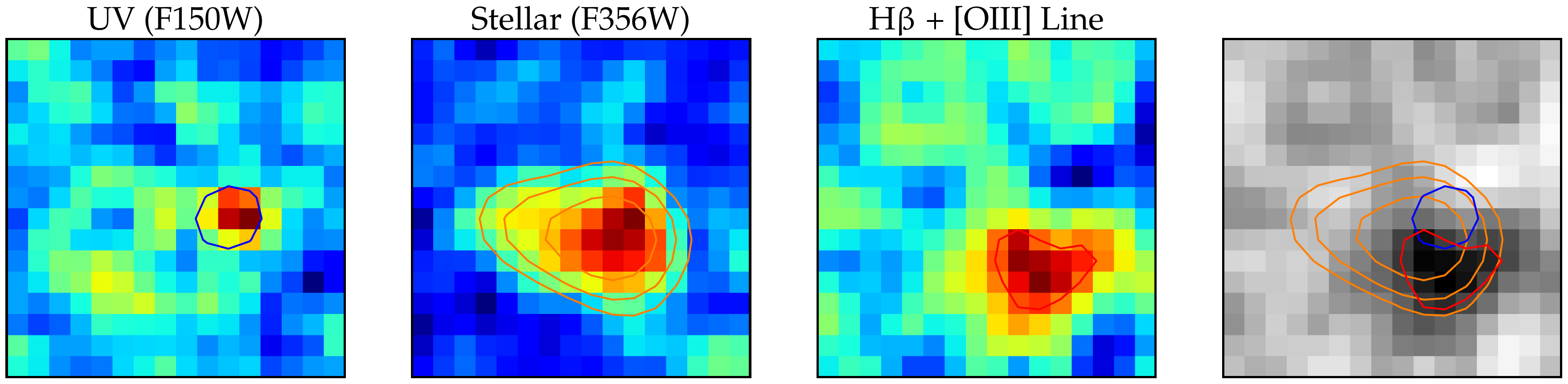}
    \includegraphics[width=168mm]{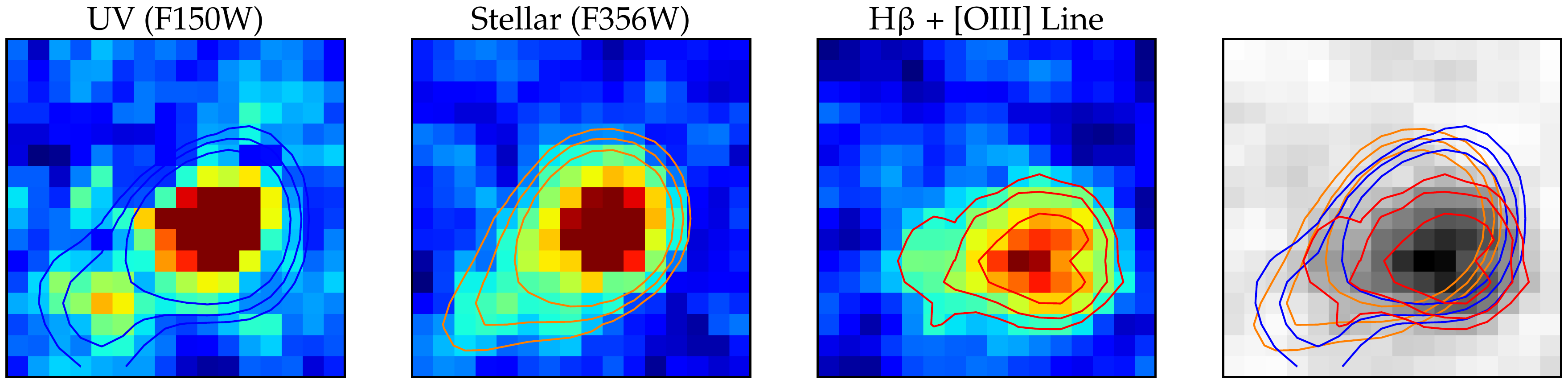}
    \includegraphics[width=168mm]{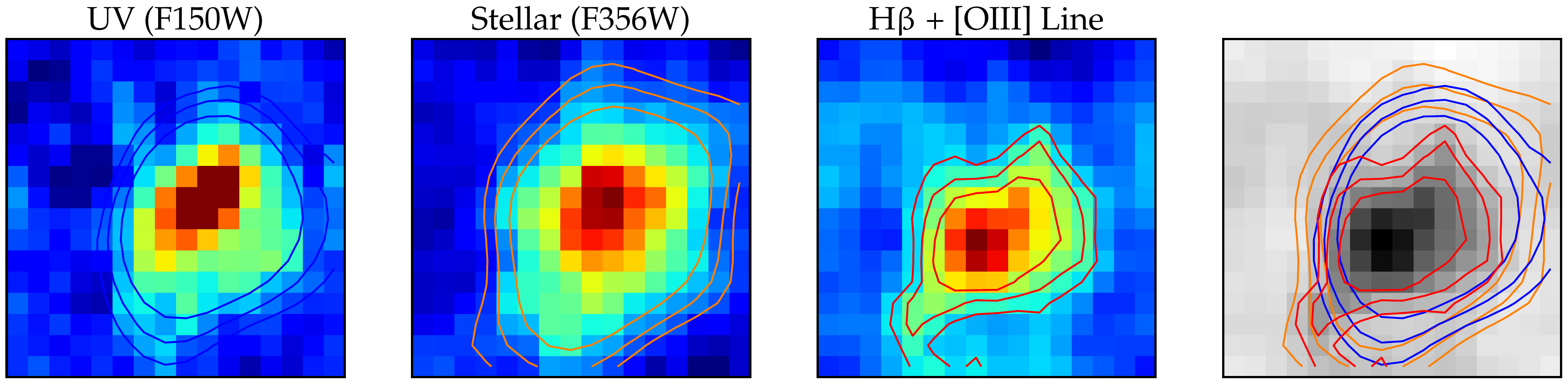}
    \includegraphics[width=168mm]{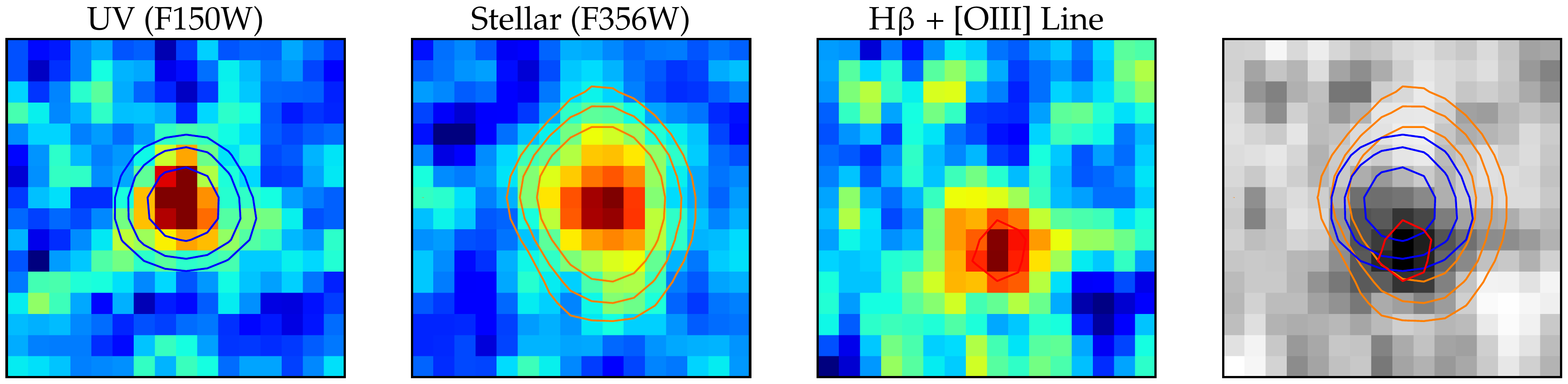}
    \caption{Image cutouts of the EELGs showing the spatial offsets between the various components. The images show UV component (F150W), stellar component (F356W; rest-frame optical), and \hb+\oiii\ emission line distributions from the left to the right panels. The contours show 5, 7, and 10 $\sigma$. In the right most panel, the emission line distribution (red contours and the underlying image) is clearly offset from the UV (blue contours) and stellar mass component (yellow contours). Note that images are the original images, but the contours are obtained from the PSF-matched images.}
    \label{fig:offset}
\end{figure*}

\begin{figure*}
    \centering
    \includegraphics[width=168mm]{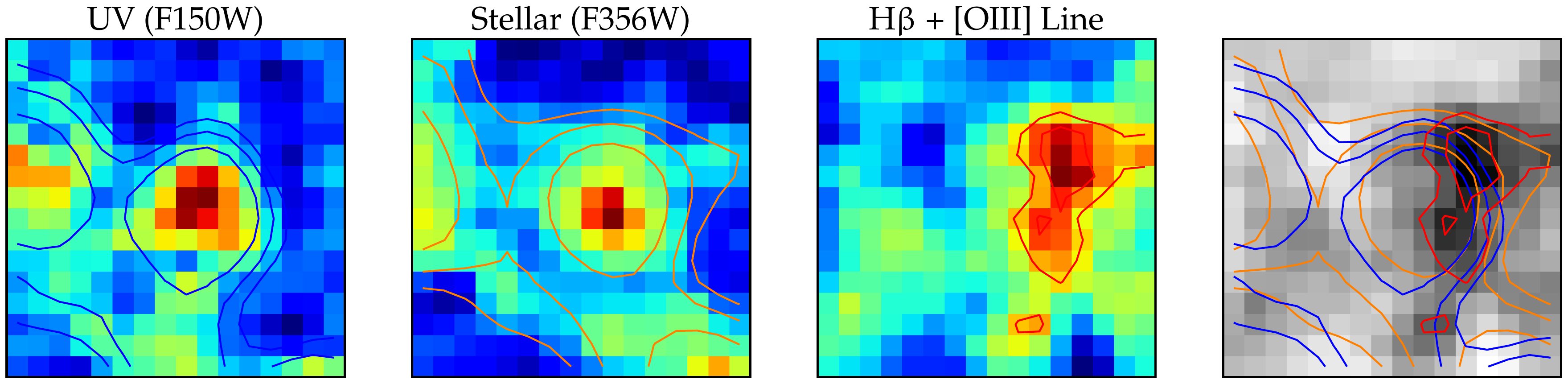}
    \includegraphics[width=168mm]{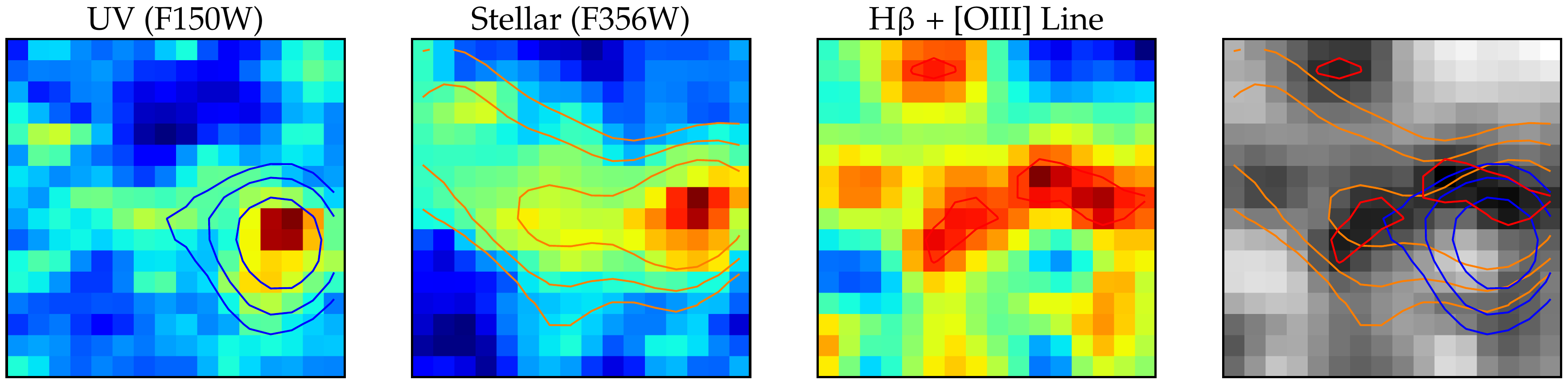}
    \includegraphics[width=168mm]{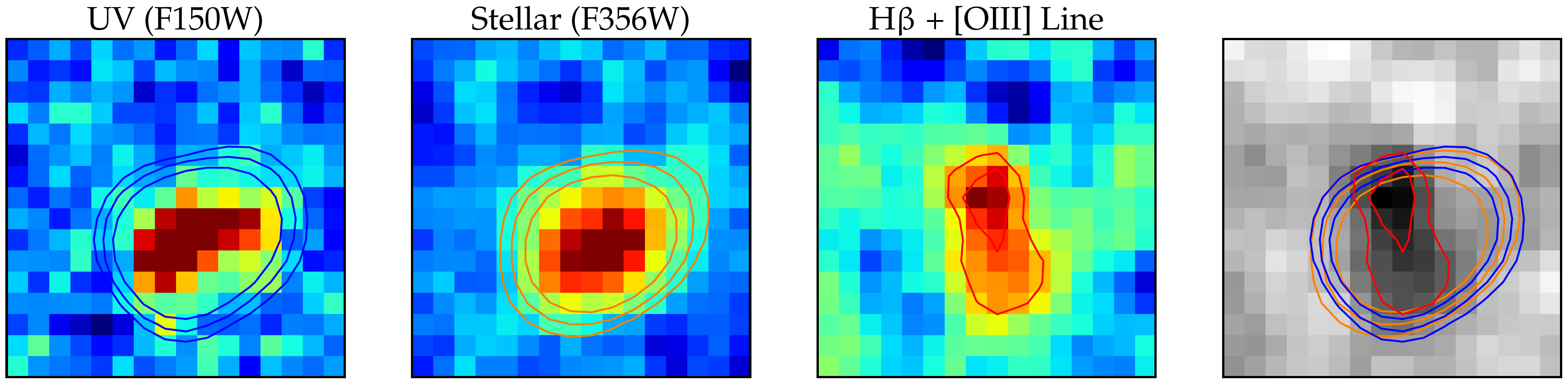}
    \includegraphics[width=168mm]{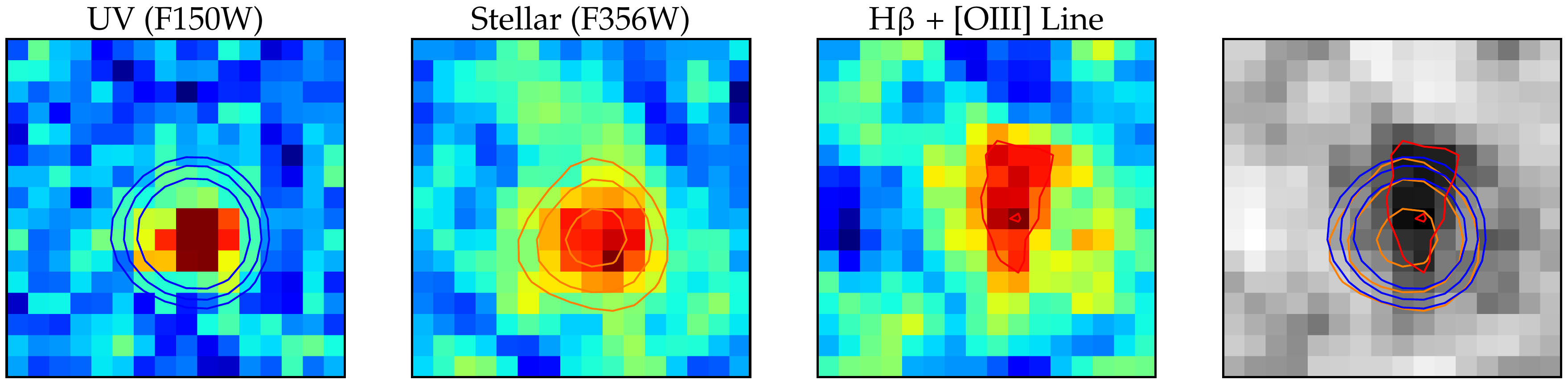}
    \includegraphics[width=168mm]{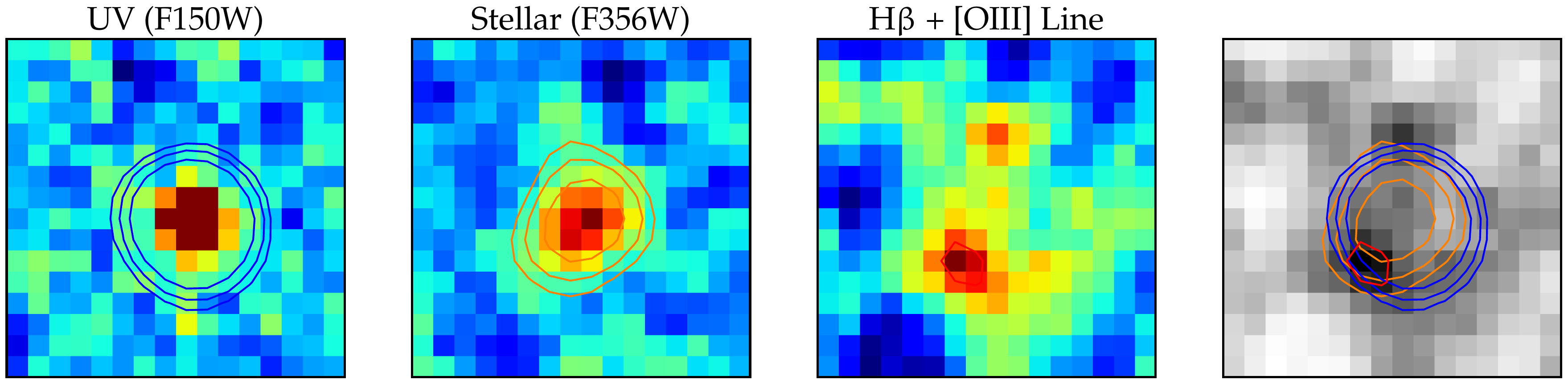}
    \caption{Image cutouts of the EELGs with the spatial offsets (ID 6 -- 10). The layout and contour specifications follow those described for Figure~\ref{fig:offset}.}
\end{figure*}

\begin{figure*}
    \centering
    \includegraphics[width=168mm]{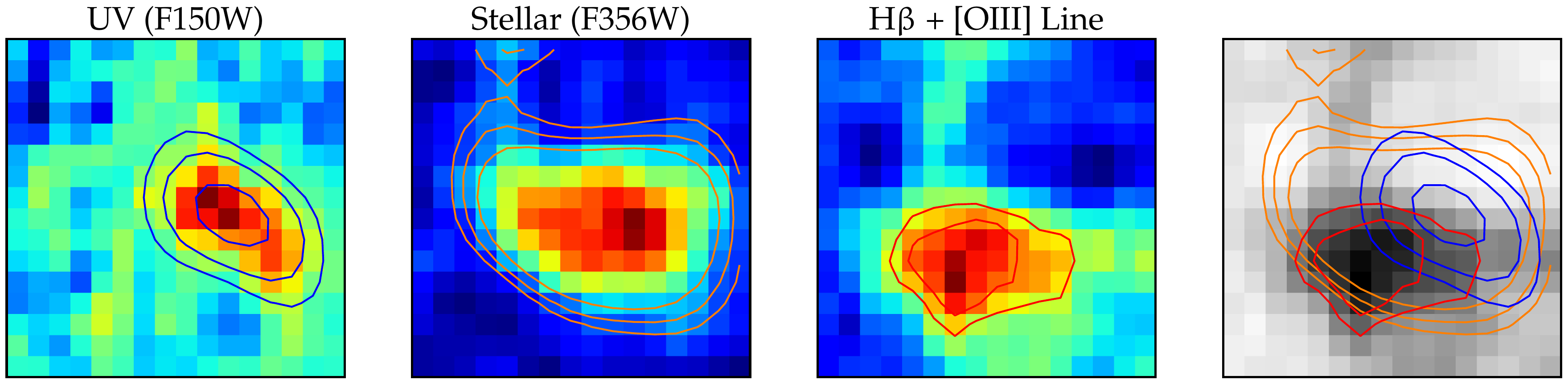}
    \includegraphics[width=168mm]{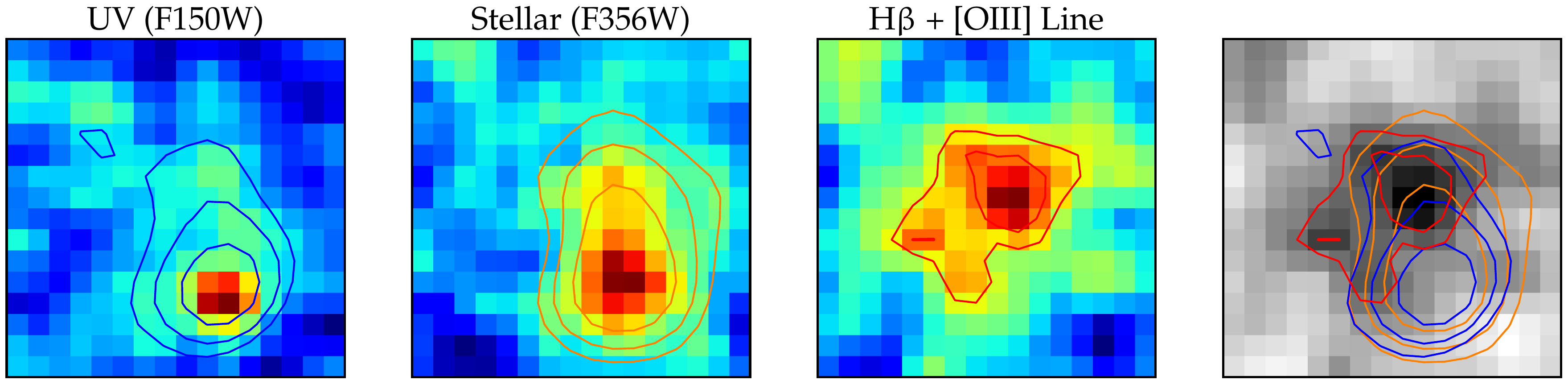}
    \includegraphics[width=168mm]{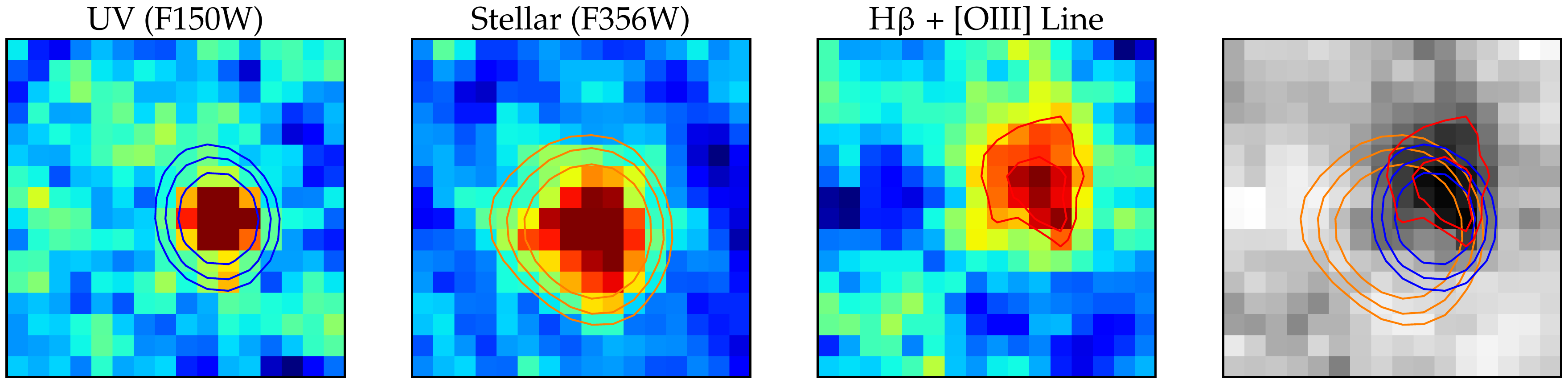}
    \caption{Image cutouts of the EELGs with the spatial offsets (ID 11 -- 13). The layout and contour specifications follow those described for Figure~\ref{fig:offset}.}
    \label{fig:offset3}
\end{figure*}

\end{document}